\documentclass[fleqn,usenatbib]{mnras}

\usepackage{newtxtext,newtxmath}

\usepackage[T1]{fontenc}
\usepackage[utf8]{inputenc}

\DeclareRobustCommand{\VAN}[3]{#2}
\let\VANthebibliography\thebibliography
\def\thebibliography{\DeclareRobustCommand{\VAN}[3]{##3}\VANthebibliography}

\usepackage{graphicx}	
\usepackage{amsmath}	
\usepackage{url}

\usepackage{subfig}
\usepackage{multirow}
\usepackage{lineno}

\newcommand{\Nu}{{\it NuSTAR}}
\newcommand{\Xm}{{\it XMM-Newton}}
\newcommand{\source}{{TXS\,1515--273}}

\title[TXS~1515--273]{First detection of VHE gamma-ray emission from TXS~1515--273, 
study of its X-ray variability and spectral energy distribution}

\author[MAGIC Collaboration]{
\parbox{\textwidth}{MAGIC Collaboration{\thanks{\textit{Send offprint requests to} MAGIC Collaboration (e-mail: \href{mailto:contact.magic@mpp.mpg.de}{\mbox{contact.magic@mpp.mpg.de}}). Corresponding authors are S. Loporchio, E.~Lindfors, and V.~Fallah~Ramazani.}}: V.~A.~Acciari$^{1}$,
S.~Ansoldi$^{2}$, 
L.~A.~Antonelli$^{3}$, 
A.~Arbet Engels$^{4}$, 
M.~Artero$^{5}$,
K.~Asano$^{6}$, 
D.~Baack$^{7}$
% ...
(affiliations are listed at the end of the paper)}
}

\date{Accepted XXX. Received YYY; in original form ZZZ}

\pubyear{2021}

\begin{document}
\label{firstpage}
\pagerange{\pageref{firstpage}--\pageref{lastpage}}
\maketitle

\begin{abstract}
We report here on the first multi-wavelength (MWL) campaign on the blazar \mbox{\source}, undertaken in 2019 and extending from radio to very-high-energy gamma rays (VHE). Up until now, this blazar had not been the subject of any detailed MWL observations. It has a rather hard photon index at GeV energies and was considered a candidate extreme high-synchrotron-peaked source. MAGIC observations resulted in the first-time detection of the source in VHE with a statistical significance of $7.6\sigma$. The average integral VHE flux of the source is $6\pm 1\%$ of the Crab nebula flux above 400 GeV. X-ray coverage was provided by \emph{Swift}-XRT, \Xm, and \Nu. The long continuous X-ray observations were separated by $\sim$9\,h, both showing clear hour scale flares. In the \Xm\ data, both the rise and decay timescales are longer in the soft X-ray than in the hard X-ray band, indicating the presence of a particle cooling regime. The X-ray variability timescales were used to constrain the size of the emission region and the strength of the magnetic field. The data allowed us to determine the synchrotron peak frequency and classify the source as a flaring high, but not extreme, synchrotron peaked object. Considering the constraints and variability patterns from the X-ray data, we model the broad-band spectral energy distribution. We applied a simple one-zone model, which could not reproduce the radio emission and the shape of the optical emission, and a two-component leptonic model with two interacting components, enabling us to reproduce the emission from radio to VHE band.
\end{abstract}



\section{Introduction}
Almost all of the extragalactic sources detected above 100 GeV are classified as Active Galactic Nuclei (AGNs): galaxies hosting a super-massive black hole in their centre whose gravitational potential energy is the ultimate source of the AGN luminosity. Up to $\sim$ 10\% of AGNs develop two narrow jets of relativistic
particles extending well outside the galaxy and emitting non-thermal radiation over the whole electromagnetic spectrum \citep{Padovani2017}. The spectra observed from jetted-AGNs depends strongly on the viewing angle of the jet with respect to the Earth, leading to their empirical classification. Jetted-AGNs with jets seen from large angles are classified as radio galaxies, while those seen  at small viewing angles ($\theta$ < 10$^{\circ}$ - 15$^{\circ}$) are known as blazars. Blazars' spectra are fully dominated by the jet emission, which can completely outshine the rest of the galaxy. They can be divided into flat spectrum radio quasars (FSRQs) and BL Lac objects (BL Lacs), depending on their optical spectra: FSRQs show strong, broad emission lines while BL Lacs display at most weak emission lines \citep{1991ApJ...374..431S,1991ApJS...76..813S}.

The broad-band spectral energy distribution (SED) of blazar emission is characterised by two distinct humps \citep{Ghisellini2017}. The first one peaks at infrared to X-ray frequencies and is commonly explained as due to synchrotron emission from ultra-relativistic electrons accelerated in the jet.  The second hump, peaking above MeV energies, is most likely due to inverse Compton (IC) scattering, possibly of the same electrons on their own synchrotron emission (synchrotron self-Compton, SSC). The presence of a sub-dominant hadronic component is also possible, as discussed in \cite{AHARONIAN2000377} and \cite{Murase_2012}.

The energy of the synchrotron peak leads to a further sub-classification of blazars. The peak frequency ranges from IR--optical to UV--soft-X bands in low, intermediate or high synchrotron-peaked sources \citep[LSP, ISP, HSP respectively, see][]{Abdo2010}. In particular, HSPs display a synchrotron emission peaking at frequencies $\nu_X \gtrsim 10^{15}$ Hz. Furthermore, evidence for objects with synchrotron peak frequency exceeding the HSP soft-X ray band was found in \cite{GHISELLINI1999} and \cite{Costamante2001}, with the most extreme high frequency-peaked blazars (EHBL) showing peaks above $10^{17}$ Hz \citep{2002A&A...384...56C, bonnoli_2015} and a hard spectrum (photon index $\leq 2$) at the \emph{Fermi} Large Area Telescope (LAT) energies, with the IC peak generally in the energy range above 100 GeV. The origin of such an extreme peak exceeding TeV energies and an explanation for the hard intrinsic spectrum at sub-TeV energies are still widely debated. This feature indicates the presence of a hard accelerated particle spectrum with most of the energy carried by the highest-energy particles.

As reported in \cite{1998Fossati}, evidence of an empirical sequence connecting blazar classes with their bolometric luminosity was found, with LSP showing higher luminosity with respect to HSP blazars. The sequence was  later revised by \cite{Ghisellini2017}, finding good agreement with the original one. It is worth mentioning that the existence of the blazar sequence has been disputed ever since it was proposed, and it is still under debate \citep[e.g.][]{2020arXiv200712661K}.

For any given blazar, only complete energy coverage, ranging from radio to TeV energies will allow for a robust study of the emission mechanisms. However, due to the variability of these objects, it is of crucial importance that these multi-wavelength (MWL) observations are performed simultaneously.

Although reported in the Second and the Third Catalog of AGNs detected by the \emph{Fermi}-LAT \citep{2LACat,3LAC}, in the  general \emph{Fermi}-LAT catalogues \citep{2fgl_2012,3fgl_2015,4FGL_2019} and in the high-energy catalogues \citep{1fhl_2013, 2fhl_2016,3fhl_2017}, the source \source\ has been very little studied, and it has never been investigated intensively in the X-rays before the observations reported in this work. An upper limit on its redshift $z <1.1$ was established in \cite{Kaur2018} with photometric methods. Recently, firm detection of spectral lines settled the redshift to z=0.1285 \citep{2020arXiv201205176G,BecerraGonzalez:2020cce}.

In 3FGL and all prior  \emph{Fermi}-LAT catalogues the source had been indeed classified as a blazar candidate of uncertain type  \citep{2017A&A...602A..86L} and only in 4FGL was it classified as a BL Lac object, with a photon index $\simeq 2$, which makes it an EHBL candidate \citep{4FGL_2019}. EHBLs are of special interest in searching for new VHE gamma-ray blazars today as they are still rare and little studied \citep[see e.g.][]{Biteau:2020prb}. Some of them have shown a hard-TeV behaviour, with IC emission peaking above  $\sim$ 10 TeV,  while others were classified as EHBL showing extreme behaviour only during flares. According to \cite{foffano2019} and \cite{Biteau:2020prb}, this may suggest the necessity for a  classification of the EHBL class into different sub-classes.

Triggered by flaring activity in the high-energy gamma-ray band (HE, 0.5 MeV $\le E \le $ 100 GeV) reported by the \emph{Fermi}-LAT \citep{Fermi_Atel}, a MWL campaign on \source\ was organised at the end of February 2019. During the flaring activity, the source was observed in different energy bands, ranging from radio to VHE gamma rays. Simultaneous or quasi-simultaneous observations were carried out by KVA, \emph{Swift}, \emph{XMM}, \Nu\ and MAGIC in order to investigate the location of the SED peaks and look for evidence of extreme behaviour during the flare. Following the observations MAGIC announced the first-time detection of the source in the VHE band \citep{Magic_atel}. In this paper, we report the results of this observing campaign.

The paper is organised as follows. In Section \ref{an_res_sect}, we describe the analysis in the different energy ranges and provide the results obtained. In Section \ref{mwl_sect} we present the studies of the MWL variability. In Section \ref{xray_variability_sect} we report on the detailed analysis of the available X-ray datasets, focusing on the short-timescale variability and the spectral evolution. In Section \ref{spectralenergydistribution_section} we describe the studies of the spectral energy distribution of \source, in particular its modelling and classification. Finally, in Section \ref{summary_sect} we present our conclusions.

\section{Analysis results}
\label{an_res_sect}
Here we present a summary of the MWL analysis performed on the acquired data.  The list of the instruments and the relative time-range of observations are provided in Table \ref{summ_table}. Further details on each dataset are given in the following sections. 
\begin{table}
\caption{Summary of the different observations performed.} 
\label{summ_table}
\begin{center}
\begin{tabular}{cccc}
\hline
\multicolumn{1}{c}{Instrument} &
\multicolumn{1}{c}{MJD start} &
\multicolumn{1}{c}{MJD stop}  \\
\hline
MAGIC &  58541.18 &  58547.21 \\              
\emph{Fermi}-LAT &  58540.72 &  58547.71  \\  
\Nu & 58544.59 & 58545.31 \\
\emph{XMM}  & 58543.95 & 58544.21 \\
\emph{Swift}-XRT & 58541.65 &  58560.71 \\
\emph{Swift}-UVOT & 58541.65 & 58560.93 \\
KVA & 58541.22 & 58718.86\\
Siena & 58542.17 & 58564.10\\
\hline
\end{tabular}
\end{center}
\end{table}  

\subsection{Very-High-Energy gamma rays}
The Major Atmospheric Gamma-ray Imaging Cherenkov telescope (MAGIC) \citep{Aleksi2016} is a stereoscopic system of two 17 m diameter imaging atmospheric Cherenkov telescopes. MAGIC telescopes performed observations of \source\ starting on 27 February 2019 (MJD 58541) up to 05 March 2019 (MJD 58547). The observations were performed at a high zenith angle, ranging from 55$^{\circ}$ to 62$^{\circ}$, under both dark time (4.8 hours) and moonlight conditions (3.3 hours), implying a high night sky background level. For this reason, we optimised the analysis chain for data taken under moonlight conditions. For further details on the MAGIC performance under moonlight, refer to \cite{AHNEN201729}. Data were analysed using the MAGIC analysis and reconstruction software MARS \citep{Zanin2013}. 
The observations led to a significant detection in the VHE range with a significance of 7.6$\sigma$. We derived the night-wise gamma-ray flux integrated above 400 GeV. The MAGIC observed flux is reported in Table \ref{magic_table}.

Since the acquired signal was not strong enough to evaluate the spectrum for each night, we combined all the data to obtain the overall spectrum. We fitted it with a power-law function, folded with the energy dispersion using the Bertero unfolding method \citep{BERTERO19891} and corrected for gamma-ray absorption by the interaction with the extra-galactic background light (EBL) using the \cite{dominguez_ebl} model. After the unfolding and the EBL correction, the MAGIC soft spectrum between 200 GeV and 900 GeV is well-described by a power-law model:

\begin{equation}
    \frac{dN}{dE} = N_0 \left(\frac{E}{E_0}\right) ^{-\Gamma},
\end{equation}
where $\Gamma = 3.11 \pm 0.32_\textrm{stat}$ is the photon index, $E_0 = 546\, \textrm{GeV}$ is the normalisation energy, selected as the decorrelation energy and $N_0 = (1.76 \pm 0.28_\textrm{stat})\times 10^{-11}\, \textrm{TeV}^{-1} \cdot \textrm{cm}^{-2} \cdot \textrm{s}^{-1}$ is the corresponding normalisation constant. The systematic uncertainties of the MAGIC telescopes are:  below 15\% on the absolute energy scale, 11\%–18\% on flux normalisation  and $\pm 0.15$ on the spectral slope \citep{Aleksi2016}. The soft spectrum in the VHE range suggests that the IC bump is likely peaking at GeV energies, as discussed in Section \ref{sect:sedmodeling}.

\subsection{High-Energy gamma rays}
The Large Area Telescope (LAT) instrument onboard the \emph{Fermi Gamma-Ray Space Telescope} satellite is a pair-conversion telescope with a precision converter-tracker and calorimeter that detects gamma rays \citep{Atwood2009}. 

Data from the \texttt{Source} event class were analysed using the Fermitools v1.2.1 and the \texttt{fermipy}\footnote{\url{https://fermipy.readthedocs.io/en/latest/}} v0.18 python package, applying standard quality cuts (\texttt{'DATA\_QUAL>0 \&\& LAT\_CONFIG==1'}) and the zenith distance cut ($<90 ^{\circ}$) to reduce the Earth limb contamination. Only events with reconstructed energy in the 300 MeV - 500 GeV range within 12$^{\circ}$ of the nominal position of the studied source were selected. A time window of 7 days between 27 February 2019 (MJD 58541) and 05 March 2019 (MJD 58547) was selected in temporal coincidence with MAGIC observations to evaluate the measured flux as a function of the time. A different time selection, from 27 February 2019 up to and including 03 March 2019 (MJD 58545) was selected for the spectral analysis, for reason that will be explained in detail in Section \ref{sect:sedmodeling}. A binned likelihood analysis was performed with 8 bins per energy decade for the selected region of interest. The instrument response function used was P8R3\_SOURCE\_V2. All of the sources in the 4FGL catalog \citep{4FGL_2019} were included in the model, along with the isotropic (\texttt{iso\_P8R3\_SOURCE\_V2\_v1}) and the Galactic (\texttt{gll\_iem\_v07}) models \citep{2016ApJS..223...26A} \footnote{\url{https://fermi.gsfc.nasa.gov/ssc/data/analysis/software/aux/4fgl/Galactic_Diffuse_Emission_Model_for_the_4FGL_Catalog_Analysis.pdf}}.

In the fitting procedure the spectral parameters of sources that are significantly detected within a radius of 5$^{\circ}$ around the source of interest were left free together with the normalisation of the diffuse components. All of the other catalog sources' parameters were fixed to the published 4FGL values. A dedicated likelihood analysis was performed for each time bin. The resulting light curve in daily bins is shown in Fig. \ref{fig:MWL_LC}, while the observed flux and the TS in each time bin are reported in Table \ref{fermi_table}.

The average spectrum obtained from \emph{Fermi}-LAT data is described by a power-law model, with the following spectral parameters: $N_0 = (20.5 \pm 7.2) \times 10^{-13} \, \textrm{MeV}^{-1}\cdot\textrm{cm}^{-2}\cdot \textrm{s}^{-1}$, $\Gamma = 2.2 \pm 0.3$ and $f_\textrm{E>300 MeV} = (4.5 \pm 1.3) \cdot 10^{-8} \textrm{cm}^{-2}\textrm{s}^{-1}$ GeV, with $E_0 = 2.3$ GeV.

\subsection{X-rays}
Following the detection of the source at VHE gamma rays, \source\ was observed at X-ray energies with the \Nu\ telescope, the \Xm\ observatory and the \textit{Neil Gehrels Swift observatory}. In this Section, we will present these observations and their outcomes. 

\subsubsection{\Xm\ observation}
\label{xmmData}
The \Xm\ observatory \citep{Jansen2001} is a space-based X-ray observatory which carries three medium spectral resolution X-ray telescopes, two Reflection Grating Spectrometers for high resolution spectroscopy \citep{den2001reflection} and a 0.3 m optical/UV imaging telescope on-board (see Section \ref{om_an}). The three X-ray telescopes at the focus of the European Photon Imaging Camera (EPIC, 0.2 -- 10 keV) are a pn-CCD operating in full frame mode \citep{Struder2001} and two multi-object spectrometer CCDs (MOS1 and MOS2) operating in small-window mode \citep{Turner2001}.

The ToO \Xm\ observation of \source\ closest to our campaign was performed at the beginning of 02 March 2019 for $\sim$25 ks of exposure time with the three X-ray telescopes. Using the \Xm\ Science Analysis System SAS version 17.0.0 and the latest available calibration files, we reduced and analysed the data by following the standard procedure explained in the SAS user guide\footnote{\url{https://xmm-tools.cosmos.esa.int/external/xmm_user_support/documentation/sas_usg/USG/}}. The source spectra were extracted from a source-centred circular region with a radius of 35 arcsec for all three instruments to have the best signal to noise ratio. The background spectra likewise were extracted from source-free regions on the same chips. We also used a time filter expression to divide the whole observation into short segments with durations of 300 sec. Then we extracted the final corresponding spectra from these segments to study the source variation over shorter time scales.

In order to estimate the Galactic column density of Hydrogen ($N_H$) in the direction of the source, we used the full data set obtained by all three X-ray instruments on board of \Xm. The spectra are binned in a way that each spectral bin contains 20 counts. The joint fit of the spectra is done using {\sc XSPEC v12.10.1s}, using photon absorption model folded with a power-law and log-parabola model, while the $N_H$ was set as a free parameter in our analysis. Moreover, the model was multiplied by a constant to account for the cross-calibration between the three instruments. The multiplicative factor was fixed to 1 for the EPIC-pn and left as free parameter in the models which are used for MOS1 and MOS2. The observed spectrum can be described well by a power-law model with photon index $\Gamma_\textrm{XMM}=2.53 \pm 0.01$ $(\chi^2/d.o.f.=2019/1882)$. We found that the log-parabola model cannot describe the observed spectrum better than the power-law model. The cross-calibration factor between the instruments was below 2\%. The value for the Galactic column density of Hydrogen from our spectral analysis is $N_H= (1.68 \pm 0.03) \times 10^{21}$\,cm$^{-2}$. The estimated value of $N_H$ is in agreement with the results presented in \citet{2013MNRAS.431..394W}. Therefore, the Galactic column density of Hydrogen is fixed to the estimated $N_H$ value in the spectral analysis with shorter time bins and for the analysis of data from other X-ray instruments. The observed flux with \Xm\, is reported in Table \ref{xmm_table}.

\subsubsection{\Nu\ observation}
\label{nustarData}

The Nuclear Spectroscopic Telescope Array (\Nu) carries two co-aligned grazing incidence X-ray telescope systems, Focal Plane Module A and B (FPMA and FPMB), operating in a wide energy range of 3--79 keV \citep{Harrison2013}. These independent CdZnTe detector units provide X-ray imaging resolution of 18$^{\prime\prime}$ (full width at half-maximum, FWHM) and spectral resolution of 400 eV (FWHM) at 10 keV. The \Nu\ ToO observation of \source\ was performed at the end of 02 March 2019 ($\sim9$\,h after the end of \Xm\ observation) for a 34 ks exposure time. The data reduction and product extraction were done using the \Nu\ Data Analysis Software {\sc nustardas v1.8.0} with a {\sc caldb} version 20180419. We performed the standard data reduction procedure explained in the \Nu\ user guide\footnote{\url{https://nustar.ssdc.asi.it/news.php\#}}. In order to extract the source spectrum, a source-centred circular region with a radius of 30 arcsec was used for both FPMA and FPMB. Likewise, the background was extracted from a source-free region with a larger radius of 60 arcsec. By creating user-selected good time intervals, we divided each exposure into 6 segments which enabled us to extract the spectra with durations of about 500--600 s each. 

We fitted both spectra collected from FPMA and FPMB simultaneously in {\sc XSPEC v12.10.1s} using a cross-calibration normalisation for each segment of observations. Due to the low count rate, we grouped all of the X-ray spectra to have at least 1 count per each energy bin. We then took the W-statistics \citep{Wachter1979} into account to do the fitting procedure. The photon absorption model folded with a power-law is used by assuming a fixed $N_H$, which was estimated in Section \ref{xmmData}. The cross-calibration factor was always below 3\%. The results of this analysis are presented in Section \ref{xray_variability_sect}. The \Nu\, observed flux is reported in Table \ref{nu_table}.

\subsubsection{\textit{Swift}-XRT observations}
\label{SwiftData}

The {\em Neil Gehrels Swift (Swift)} satellite \citep{gehrels04} carried out 8 observations of \source\ between 30 September 2014 and 18 March 2019. The observations were performed with all three instruments on board: the X-ray Telescope \citep[XRT;][0.2--10.0 keV]{burrows05}, the Ultraviolet/Optical Telescope \citep[UVOT;][170--600 nm]{roming05} (see Section \ref{uvotdata}), and the Burst Alert Telescope \citep[BAT;][15--150 keV]{barthelmy05}. The hard X-ray flux of this source turned out to be below the sensitivity of the BAT instrument for such short exposures and therefore the data from this instrument will not be used. Moreover, the source was not included in the \emph{Swift}-BAT 70-month hard X-ray catalogue \citep{oh18}.

The multi-epoch event lists of the \textit{Swift}-XRT data were downloaded from the publicly available \textit{Swift}-XRT Instrument Log \footnote{\url{https://heasarc.gsfc.nasa.gov/W3Browse/swift/swiftxrlog.html}}. These observations were carried out in photon-counting mode. Following the standard \textit{Swift}-XRT analysis procedure described by \citet{2009MNRAS.397.1177E}, the data were processed using the configuration described by \citet{2017A&A...608A..68F} for blazars, assuming a photon absorption model folded with a power-law model and fixed $N_H$ as estimated in Section \ref{xmmData}. In Table \ref{XRT_table} we provide the results obtained from fitting the {\em Swift}-XRT spectrum. 

Two \emph{Swift}-XRT snapshots (MJD 58544.84 and 58544.97), simultaneous with the \Nu\ data, are combined with each other. A joint-fit was performed using these  XRT data set and the full data obtained by \Nu. The observed spectra can be described by a photon absorption model (assuming fixed $N_H$) folded with a broken power-law model $(\chi^2/d.o.f.=405/378)$. The spectrum shows a break at $3.34\pm 0.34$\,keV. The photon indexes before and after the break energy are $\Gamma_1=2.10\pm0.11$  and  $\Gamma_2=2.72\pm0.03$, respectively. The cross-calibration factor between \textit{Swift}-XRT and \Nu\ data was 15\%. 

\subsection{Optical and UV}

\subsubsection{{\it Swift}-UVOT observations}
\label{uvotdata}

During the {\em Swift} pointings, the UVOT instrument observed \source\ in all of its optical ($v$, $b$ and $u$) and UV ($w1$, $m2$ and $w2$) photometric bands \citep{poole08,breeveld10}. We analysed the data using the \texttt{uvotsource} task included in the \texttt{HEAsoft} package (v6.28) with the 20201026 release of the Swift/UVOTA CALDB. Source counts were extracted from a circular region of 5 arcsec radius centred on the source, while background counts were derived from a circular region of 20 arcsec radius in a nearby source-free region. The observed magnitudes are reported in Table~\ref{uvot_table}.

The UVOT flux densities were corrected for Galactic extinction using the E(B--V) value of 0.209 from \citet{schlafly11} and the extinction laws from \citet{cardelli89}. From the V-band fluxes, we subtracted the host galaxy contribution, which was estimated assuming a host galaxy absolute magnitude $M_R=-22.8$, galaxy colour V-R = 0.8 \citep{fukugita95}, and $z$ = 0.1285. Within the aperture of five arcsec, its contribution is 0.13\,mJy. We corrected only the V-band because the host galaxy contribution to other UVOT bands is negligible.  We note that the magnitudes in Table~\ref{uvot_table} are the observed ones, i.e. corrections for neither the host galaxy contribution nor the Galactic extinction have been performed.

\subsubsection{{\it XMM-OM} observations}
\label{om_an}
The Optical Monitor (OM) observed the source in the $b$, $u$ and $w1$ filters in imaging mode. The total exposure times of the imaging observations are approximately: 9400 s, 4700 s, and 9400 s. The data were processed using the SAS task \texttt{omichain}. The count rate is converted to flux using the conversion factors given in the SAS watchout dedicated page \footnote{\url{https://www.cosmos.esa.int/web/xmm-newton/sas-watchout-uvflux}}. The observed magnitudes with \emph{XMM-OM} are reported in Table \ref{om_table}.
These magnitudes were then corrected for extinction using the same E(B--V) value and extinction laws as for the UVOT data. 

\subsubsection{KVA}
\label{KVA}
The optical R-band observations were performed using the 35 cm telescope attached to the 60 cm KVA telescope located at La Palma. The data analysis was performed  using  the  standard  procedures  with the semi-automatic pipeline developed in Tuorla \citep{nilsson_kva}. As the source is not part of the Tuorla Blazar monitoring program\footnote{\url{http://users.utu.fi/kani/1m/index.html}}, a proper calibration was required in order to perform differential photometry. A comparison star and a control star were selected among known stars in the same field of view and were calibrated with respect to stars in the field of other targets observed on the same night under photometric conditions. 

The magnitudes were corrected for Galactic extinction using the same E(B-V) value as for correcting the UVOT data. The host galaxy flux was estimated in the same way as for the UVOT, resulting in a contribution of 0.303\,mJy within the used aperture of 5 arcseconds.

Observations of the source were performed for several months after the flaring state to estimate the flux during quiescent periods. The overall light curve is reported in Fig. \ref{fig:longterm_lc}, while the observed magnitudes (without host galaxy subtraction or Galactic extinction corrections) are reported in Table \ref{kva_table}.

\subsubsection{Siena}

The Astronomical Observatory of the University of Siena observed \source\ in the context of a program focused on optical photometry of blazars in support of MAGIC. The instrumentation consists of a remotely operated 30 cm Maksutov-Cassegrain telescope installed on a Comec 10 micron GM2000-QCI equatorial mount. The detector is a Sbig STL-6303 camera equipped with a 3072 x 2048 pixels KAF-6303E sensor; the filter wheel hosts a set of Johnson-Cousins BVRI filters. Multiple 300 s images of \source\ were acquired in the R band at each visit. Observations were always performed closely around culmination given the southern declination of the source. After standard dark current subtraction and flat-fielding, images for each visit were averaged and aperture photometry was performed on the average frame by means of the MaximDL software package. The choice of reference and control stars was consistent with the one for the KVA data. The obtained magnitudes reported in Table \ref{siena_table} were corrected for Galactic extinction and the host galaxy magnitude, which results in a contribution of 0.37\,mJy within the used aperture of 7 arcseconds, was subtracted as was done for the KVA data. Table \ref{siena_table} reports the magnitudes without these corrections.

\subsection{Very Long Baseline Array}
\source\ has been observed six times with the Very Long Baseline Array (VLBA) Experiment \citep{Lister_2019} at 15\,GHz as part of the Monitoring Of Jets in Active galactic nuclei with VLBA Experiments (MOJAVE) 
program{\footnote{\url{https://www.physics.purdue.edu/MOJAVE/}}}. It was observed three times in 2017 (28 January, 25 May and 17 June), once in 2018 (31 May) and twice in 2019 (13 June and 19 July). We retrieve the fully-calibrated MOJAVE data sets and we produced images in Stokes' I, Q, and U using the task IMAGR in AIPS. We then combined images in Stokes Q and U and produced polarisation intensity images and the associated error maps.
We fitted the visibility data with circular Gaussian components at each epoch using the model-fit option in DIFMAP. Model-fitting the visibility reveals the core and two quasi-stationary components: the first one at 0.15-0.2 milliarcseconds from the core and the second one at 1.5 milliarcseconds (both with a nearly constant flux). From the data, there is no evidence for a new jet component emerging from the core, but the cadence of the observations is not optimal for detecting new components.
The core flux was almost constant within the error bars, varying between $81\pm6$ (June 2019) and $99\pm7$ mJy (May 2017), in the first five epochs, while in the last epoch (July 2019) there is marginal hint of increased flux ($109\pm8$ mJy).

The source was not polarised in 2017 with an upper limit to the polarised flux density $\leq$0.2 mJy and to the fractional polarisation $\leq$0.15\%. In 2018 May the polarised flux increased to 0.8$\pm$0.1 mJy, corresponding to a fractional polarisation of 0.5\%. In 2019 the polarised flux increased again to 1.5$\pm$0.2 mJy (about 1.5\%) and 1.0$\pm$0.1 mJy (0.8\%) in June and July, respectively. However, the sparse time coverage does not allow us to set any robust connection between the polarised emission and the MWL activity.

\section{Multi-Wavelength variability}
\label{mwl_sect}
\begin{figure*}
 \includegraphics[width=2\columnwidth]{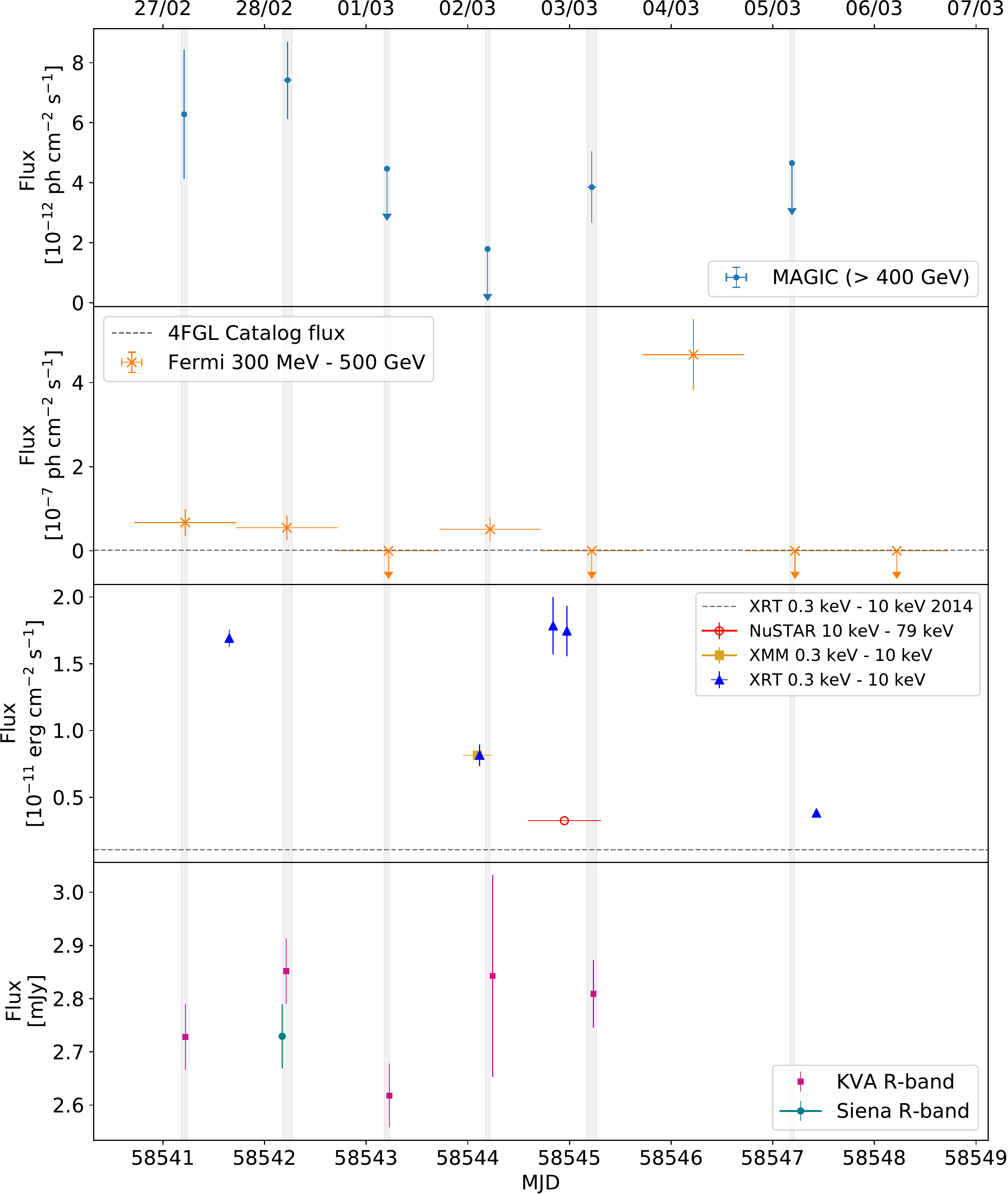}
 \caption{MWL light curve of \source\ from 27 February 2019 to 06 March 2019. From top to bottom: VHE gamma rays by MAGIC, HE gamma rays by \emph{Fermi}-LAT, X-rays by \emph{Swift}-XRT (blue triangles), \Nu\ (red open circles) and \Xm\ (yellow squares) in different energy bands and R-band by KVA and Siena observatory. 95\% confidence upper limits are indicated as downward arrows in VHE gamma rays where the flux is compatible with zero as well as in the HE band for each time bin where the test statistic (TS) value for the source was found to be smaller than 9. The individual light curves are daily binned. Dashed horizontal grey lines indicate the level of detected flux during 2014 \emph{Swift}-XRT observations and the reference flux level in the 4FGL catalog. The grey vertical bands are drawn to highlight the MAGIC observation time slots. Magnitudes for optical data were corrected for Galactic extinction, and the host galaxy contribution was subtracted (details in the text).
 }
 \label{fig:MWL_LC}
\end{figure*}

\begin{figure}
\includegraphics[width=1.1\columnwidth]{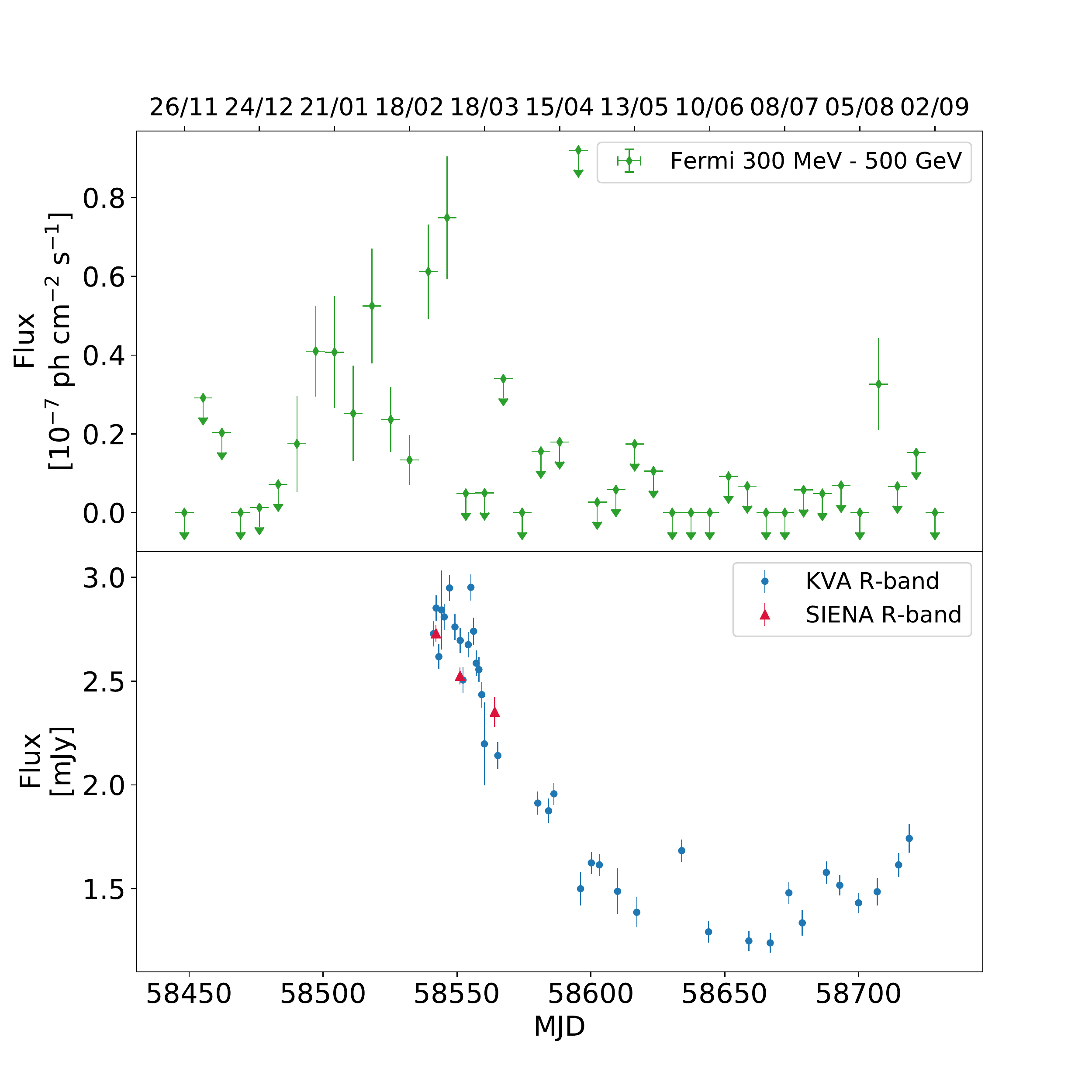}
\caption{ \label{fig:longterm_lc} HE gamma-ray (top, green diamonds) and R-band (bottom, KVA in blue circles and Siena observatory in red triangles) light curves of \source. \emph{Fermi}-LAT light curve is shown in weekly bins, with 95\% confidence UL indicated. The light curve starts from a few months before the flare period up to August 2019. The large flare can be seen during our observations. The optical light curve has daily bins, starting from the observations carried out during the flare.
}
\end{figure}

The light curves of \source\ in the different wavelengths from radio to VHE gamma rays are reported in Fig. \ref{fig:MWL_LC}. The MWL light curves include all observations from MJD 58541 to MJD 58548. For comparison purposes, we show dashed grey lines in the \emph{Fermi}-LAT and \emph{Swift}-XRT light curves representing the reference value flux from the 4FGL catalog \citep{4FGL_2019} and a previous detection from \emph{Swift}-XRT. From the reference level of the 4FGL catalog and a previous observation in the X-rays, it is evident that the source was in a high state in the X-ray and the HE band. The long term optical and HE gamma-ray light curves shown in Fig. \ref{fig:longterm_lc} also indicate clearly that the observed flux in those energy ranges was higher than usual. A significant increase of activity was also observed by {\em Swift}-UVOT between 2014 and 2019 with all optical and UV filters (see Table \ref{uvot_table}).

The \emph{Fermi}-LAT light curve shows a significantly higher flux in HE during the night of March 04, 2019, when the MAGIC observations in VHE gamma rays were prevented by bad weather. The \emph{Fermi}-LAT SED was evaluated up to and including the night of March 03, 2019, to exclude this high state for which we did not have MAGIC data (see Fig.~\ref{fig:MWL_LC} and Section \ref{sect:sedmodeling}).

We also searched for intra-night variability in different bands. 
Intra-night variability was detected in X-ray observations on 02 March 2019 with both \Xm\ and \Nu. The longest continuous MAGIC observations were $\sim2$ hours, but due to lack of statistics we could not investigate further for intra-night variability.
In the \emph{Fermi}-LAT band the source is too weak to detect hour-scale variability as found in X-ray band. In optical band the variability in general has a rather small amplitude. 
We remark, however, that the \Xm\ and \Nu\ observations, despite being very close in time, were not overlapping: as a result, different variability timescales are not unexpected, as indeed was found (see Section \ref{xray_variability_sect}). Consequently, we also decided to model the SED separately for the \Xm\ and \Nu\ epochs, as we will discuss in detail in Section \ref{sect:sedmodeling}. 

\section{Short-time X-ray variability}
\label{xray_variability_sect}
Our observing campaign had particularly good coverage in X-rays, including \Xm, \emph{Swift}-XRT and \Nu. The long exposures of \Xm\ and \Nu\ observations allow us to investigate in detail the short-time scale variability of the source in X-rays. 
\subsection{Flare timescales}
\label{timescale_fit}
\begin{figure}
\centering
\subfloat[][ \Xm\ 0.3 -- 3 keV 5-minute bins light curve for \source.]
{\includegraphics[width=\columnwidth]{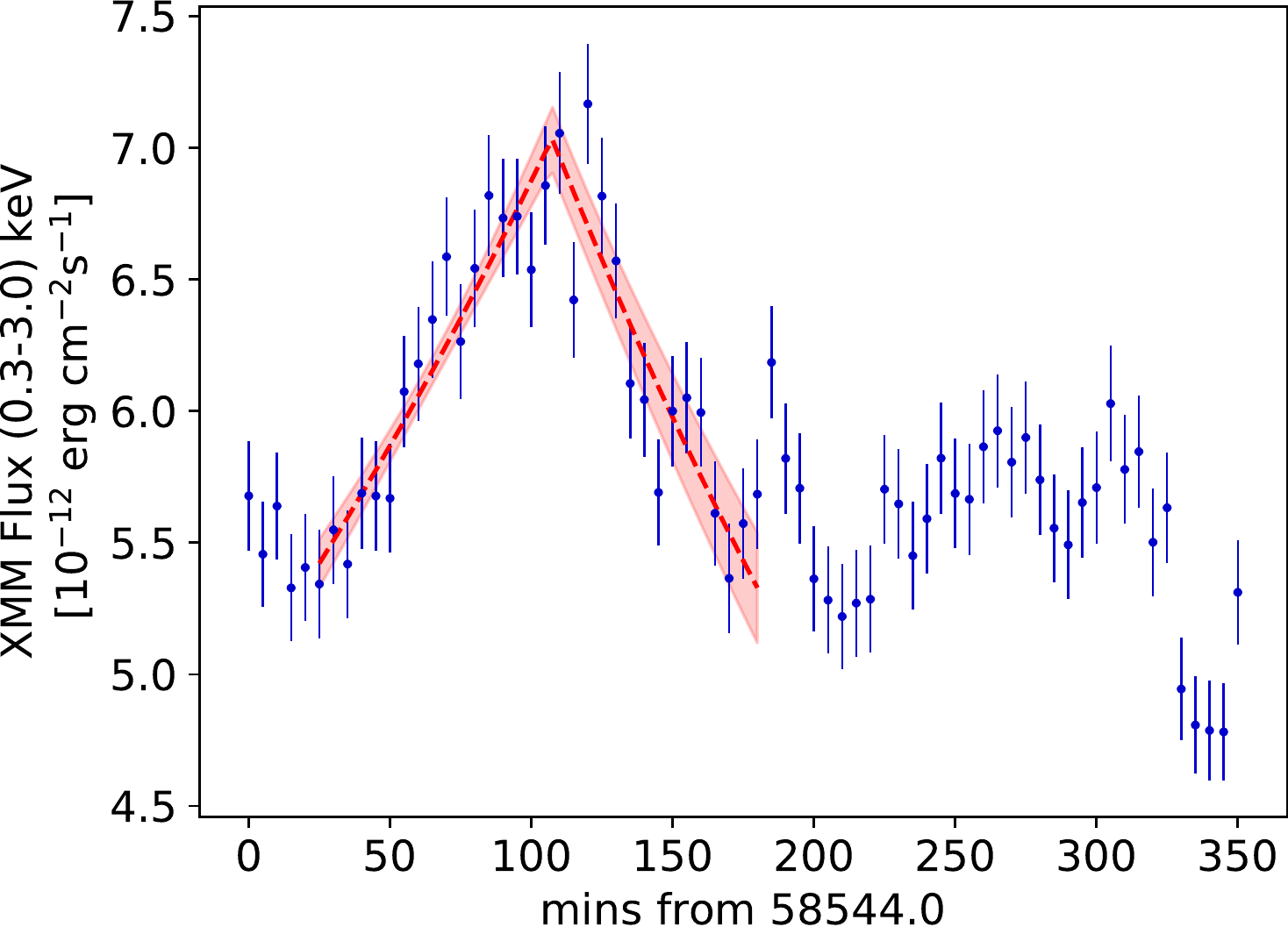}}\\
\subfloat[][ \Xm\ 3 -- 10 keV 5-minute bins light curve for \source.]
{\includegraphics[width=\columnwidth]{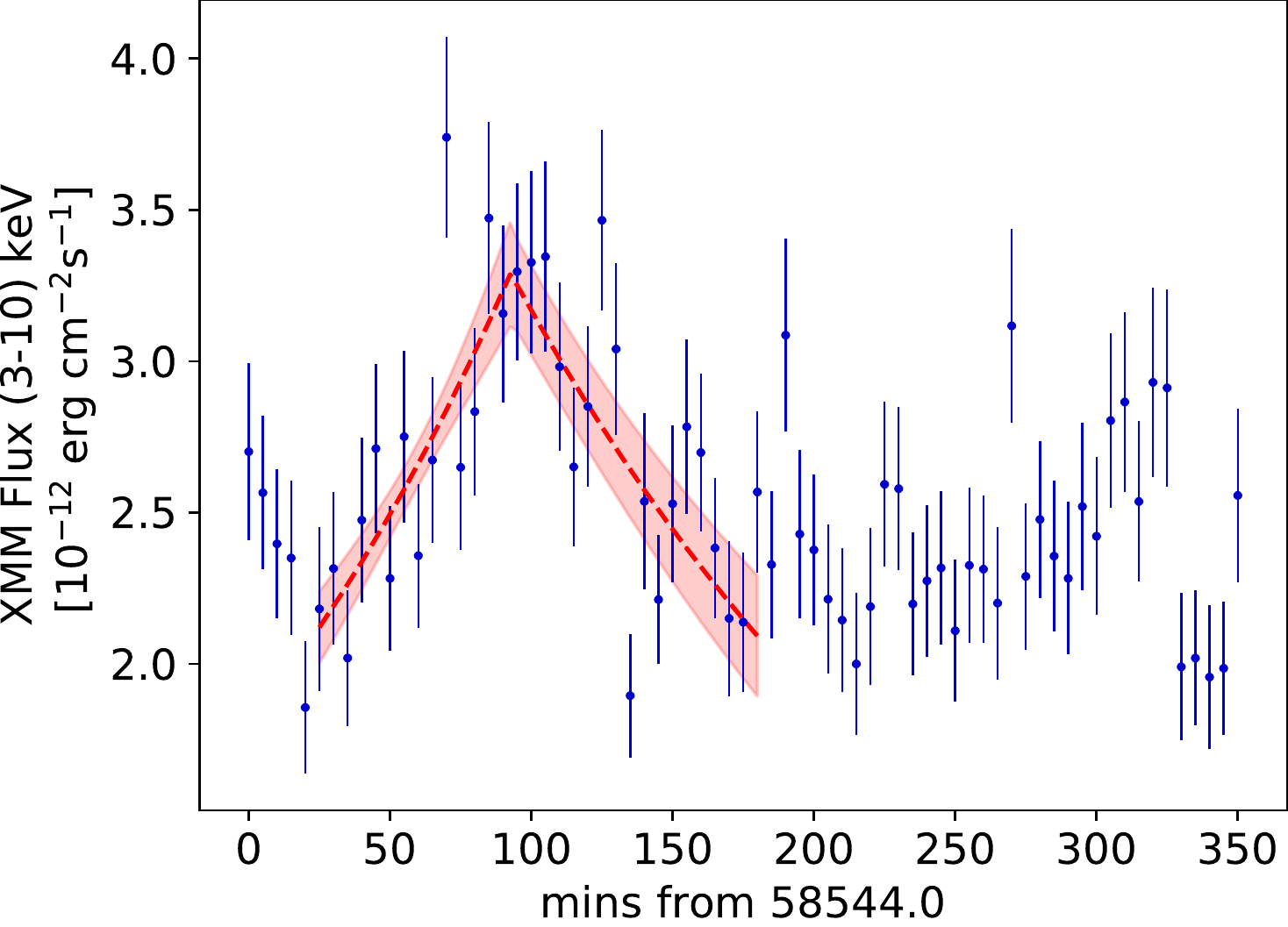}}
\caption{\Xm\ light curves for \source\ for soft (top) and hard (bottom) energy band. The dashed curves superimposed on the data represents the exponential function used to fit the peak, the rise and decay time of the bursts. The red shaded areas represent the uncertainties of the fit.}
\label{fig:exp_fit_xmm}
\end{figure}

\begin{figure}
\centering
\subfloat[][\Nu\ 3 -- 10 keV 10-minute bins light curve for \source.]
 {\includegraphics[width=\columnwidth]{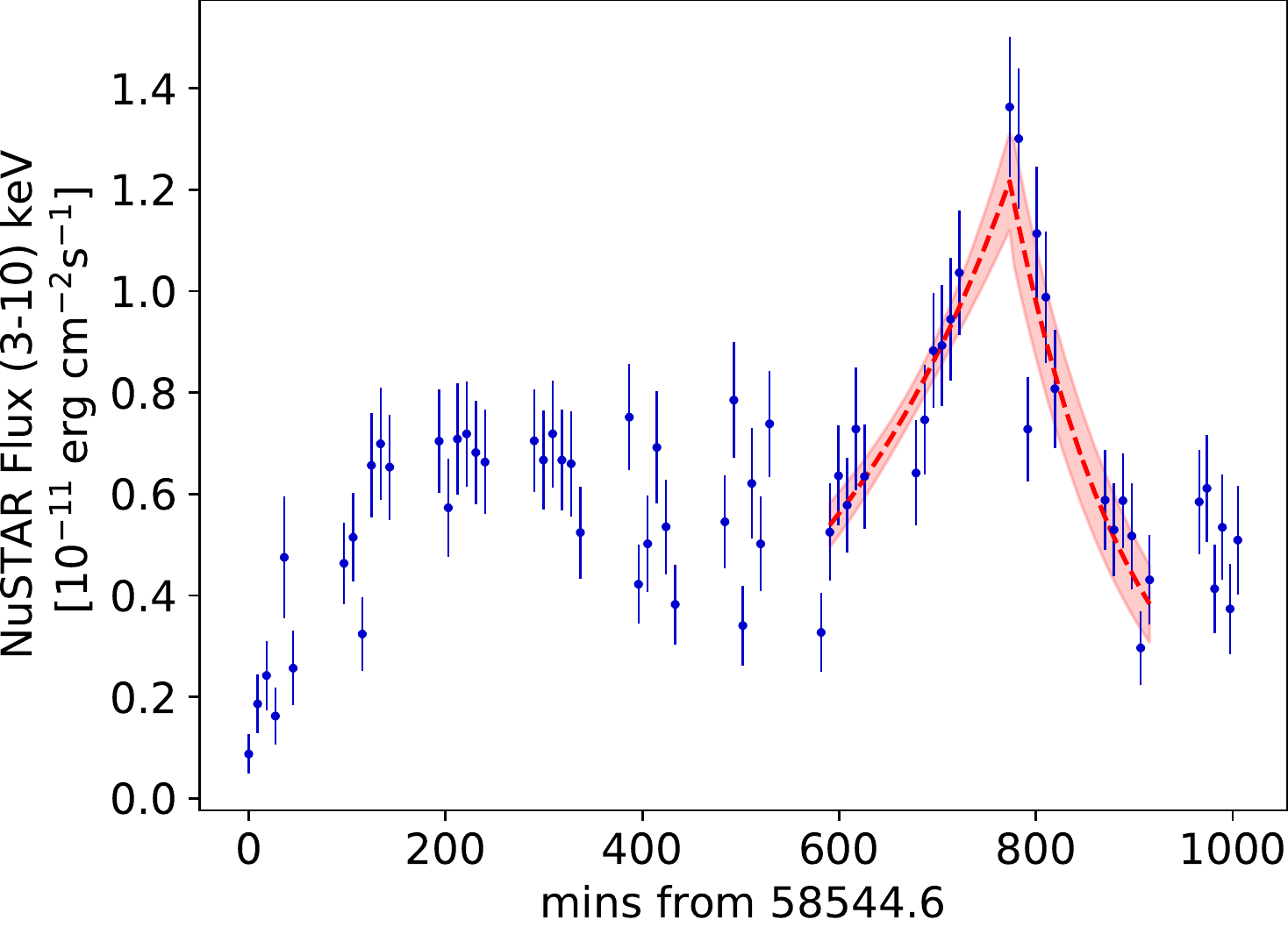}}\\
 \subfloat[][\Nu\ 10 -- 79 keV 10-minute bins light curve for \source.]
{\includegraphics[width=\columnwidth]{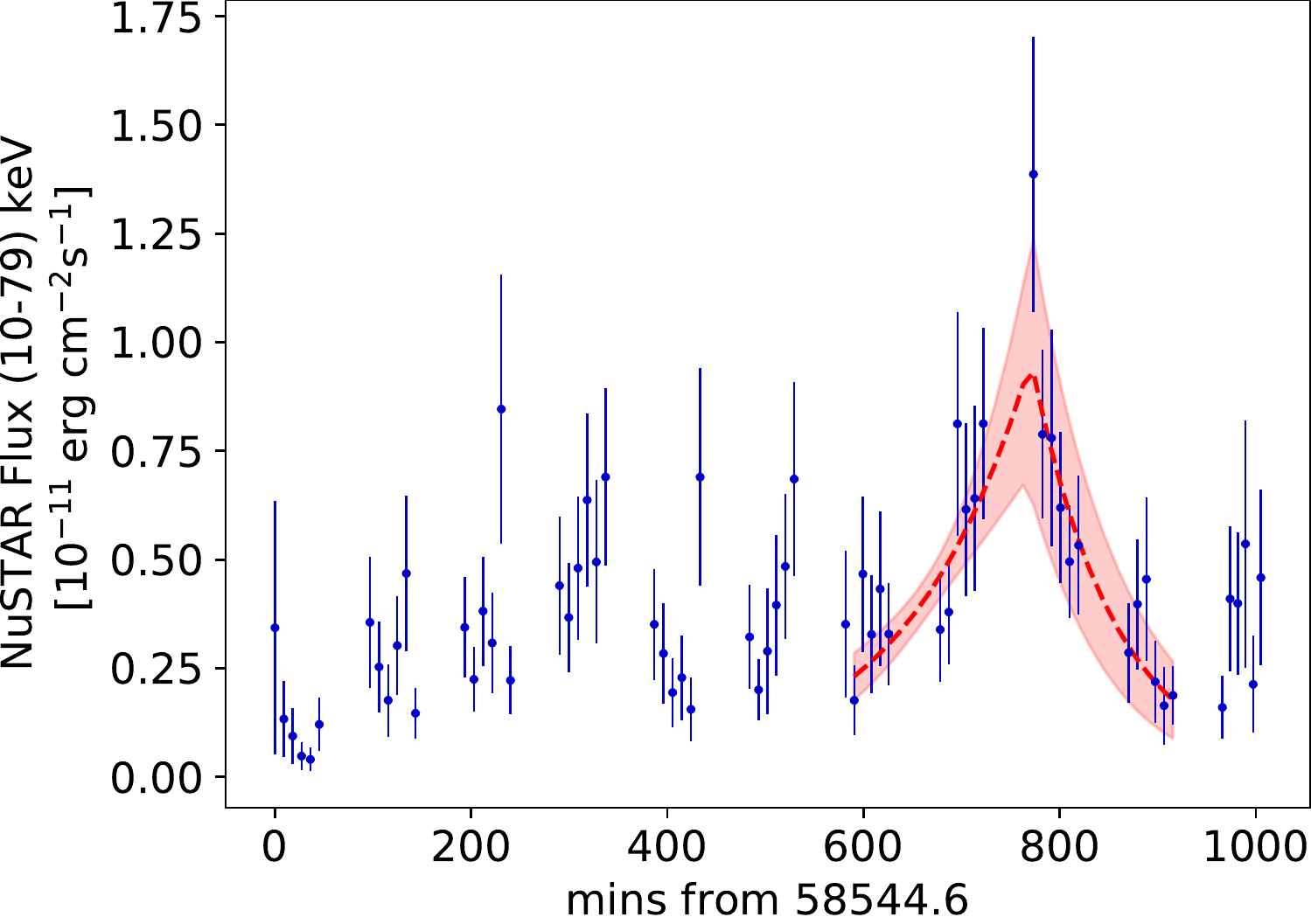}}
\caption{\Nu\ light curves for \source\ for soft (top) and hard (bottom) energy band. The dashed curves superimposed on the data represents the exponential function used to fit the peak, the rise and decay time of the bursts. The red shaded areas represent the uncertainties of the fit.}
\label{fig:exp_fit_nu}
\end{figure}

The light curves showed multiple flares (see Fig.\ref{fig:exp_fit_xmm} and \ref{fig:exp_fit_nu}) and in the following we investigate the details of this variability and use it to constrain physical parameters of the emission region.
    
To constrain the variability timescale, an exponential function in the form
\begin{equation}
F(t) = A \cdot \begin{cases}
\exp [(t - t_\textrm{peak})/ t_\textrm{rise}] \qquad  \textrm{if} \qquad t < t_\textrm{peak}\\
\exp[(t_\textrm{peak}-t) / t_\textrm{decay}] \qquad  \textrm{if} \qquad t > t_\textrm{peak}
\end{cases}
\end{equation}
was used to fit the peak time, the rise and fall profiles of the bursts visible in the light curves. In both epochs, we fitted the soft and the hard energy band light curves independently, as shown in Figs \ref{fig:exp_fit_xmm} and \ref{fig:exp_fit_nu}.
The results of the fit are reported in Table \ref{tab:burstfit}. We cross-checked the results of our fit with different time binning of the light curves, i.e. 5, 10 and 15 minutes for \Xm\ and $\sim10$ and $\sim20$ minutes for \Nu. As a confirmation of the consistency of the results, no significant difference was found for the different binning.

We also tried several combinations for start and stop bins. First we selected a reasonable range for them to vary, performing the fit of the flare profile with different combinations of the range edges. We always kept the same fit range for soft and hard light curves.
Finally, we picked the range that gave the best agreement with experimental data, i.e. with the minimum $\chi^2$ value. The parameter $t_\textrm{peak}$ was always set free to vary during the fitting procedure.

As shown in Figs \ref{fig:exp_fit_xmm} and \ref{fig:exp_fit_nu}, rapid variations on the time scale of the order of hours were detected in the two epochs. 
The derived timescales are not anomalous compared to X-ray flares seen in other blazars \citep{2020arXiv200108678M,2020A&A...638A..14M}. 

The flare profiles in all bands are symmetric, i.e. within the derived errorbars the rise and decay timescales of the flares are not significantly different. The rise times of both the observed X-ray flares are longer in the low-energy band than in the high-energy band. The difference is significant both in the \Xm\ and \Nu\ results, with a confidence level higher than 99\% and 90\% for the two datasets respectively. In standard acceleration scenarios, low-energy electrons are supposed to be accelerated faster than high-energy ones. If the rise-time of the flare is dominated by the acceleration time-scale, then the rise-time in the lower energy band is expected to be shorter than the one in the high-energy band. However, higher energy electrons cool faster than lower energy electrons, and therefore in a cooling dominated regime one would indeed expect the higher energies to rise faster \citep[e.g.][]{kirk1998} as seen in our data. 
This should also result in the high-energy flux peaking earlier than the lower energy flux, as seen in \Xm\ data. In \Nu\ data this is not evident, which is in line with lower statistical significance in difference of rise times between the two bands as well.

\begin{table}
\setlength{\tabcolsep}{0.41em}
\caption{Results of the fit of the burst profile for the \Xm\ and the \Nu\ epoch. The starting time corresponds to MJD 58543.0 for \Xm\ and to MJD 58544.6 for \Nu.}
\label{tab:burstfit}
\begin{center}
\begin{tabular}{ccccc}
\hline
\multicolumn{1}{c}{Instrument and energy} &
\multicolumn{1}{c}{$t_\textrm{rise}$} &
\multicolumn{1}{c}{$t_\textrm{peak}$} &
\multicolumn{1}{c}{$t_\textrm{decay}$} &
\multicolumn{1}{c}{$t_\textrm{peak}$}\\
 & (min) & (min) & (min) & (MJD)\\
\hline
\emph{XMM} (0.3-3 keV) & $316\pm 34$ & $107 \pm 3$ & $262 \pm 32$ & 58544.06\\
\emph{XMM} (3-10 keV) & $159 \pm 29$ & $93 \pm 6$ & $193 \pm 37$ & 58544.06\\
\Nu\ (3-10 keV) & $225 \pm 42 $ & $ 773 \pm 10$ & $123 \pm 17$ & 58545.14\\
\Nu\ (10-79 keV) & $127 \pm 41 $ & $ 771 \pm 16$ & $86 \pm 18$ & 58545.14\\
\hline
\end{tabular}
\end{center}
\end{table}

\subsection{Spectral evolution}
\label{spectral_ev}
We investigated the dependence of the photon index on the flux in the full energy range (0.3 - 10 keV for \Xm, 3 - 79 keV for \Nu). The \Xm\ photon index showed little variations, ranging between $2.35 \pm 0.05$ and $2.76 \pm 0.06$ (using 5-minute binning within the full time range of observation), while the \Nu\ photon index showed a wider dynamical range, varying between $2.01 \pm 0.26$ and $4.33 \pm 0.71$ (using 10-minute binning within the full time range of observation). In both of the epochs studied, this dependency showed a more complicated behaviour than simple spectral hardening. We also examined the relation between the integral flux and the hardness ratio, but no clear correlation between these quantities was found in our data sample.

\begin{figure}
\centering
\subfloat[][\Xm\ photon index versus integral (0.3 - 10 keV) flux for \source\ during the short flare observed (MJD 58544.00 - 58544.08, i.e. 25 - 125 minutes after the beginning of the observations).]
{\includegraphics[width=\columnwidth]{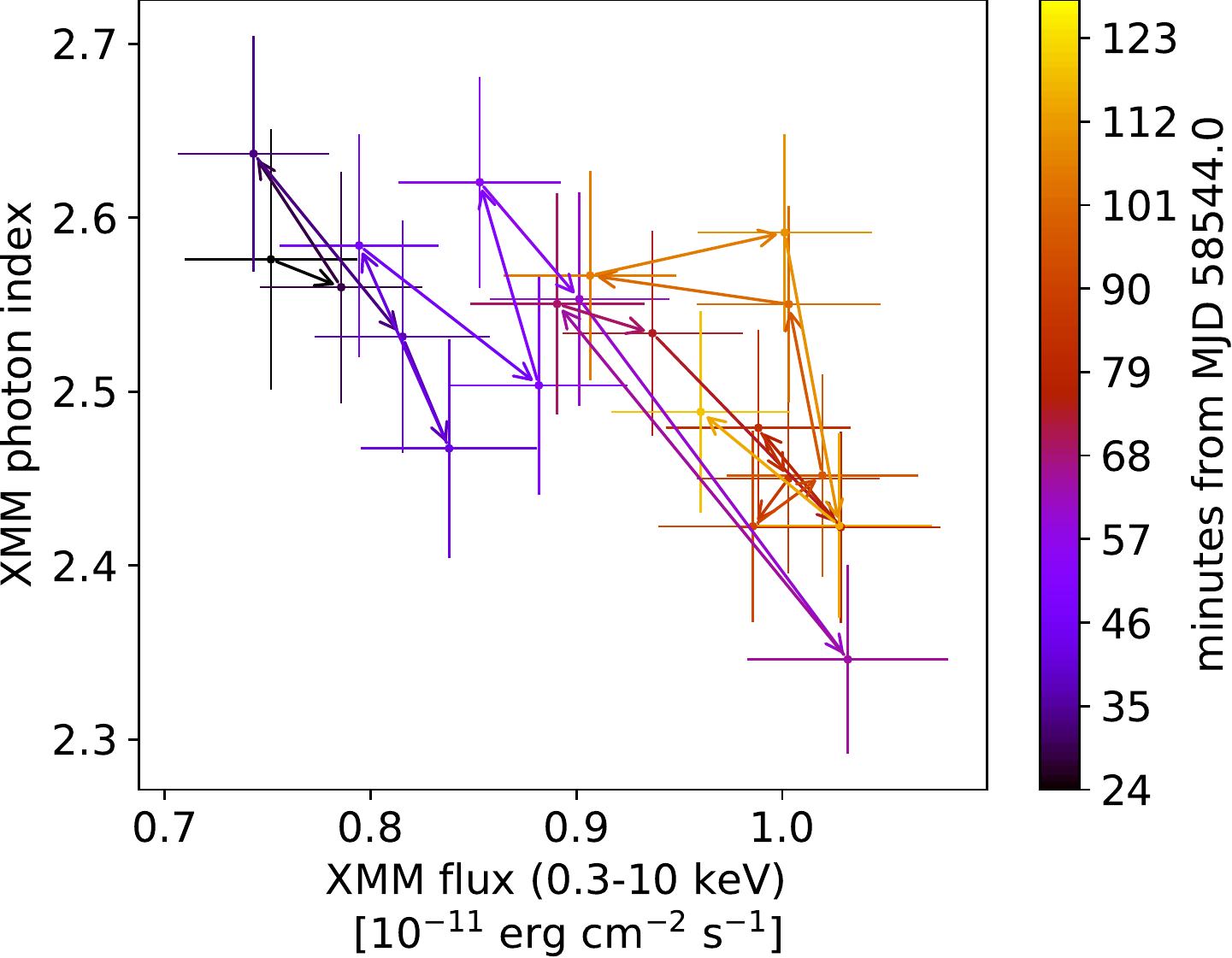}}\\
\subfloat[][\Nu\ photon index versus integral (3 - 79 keV) flux for \source\ during the short flare observed (MJD 58545.08 - 58545.17, i.e. 670 - 810 minutes after the beginning of the observations).]
{\includegraphics[width=\columnwidth]{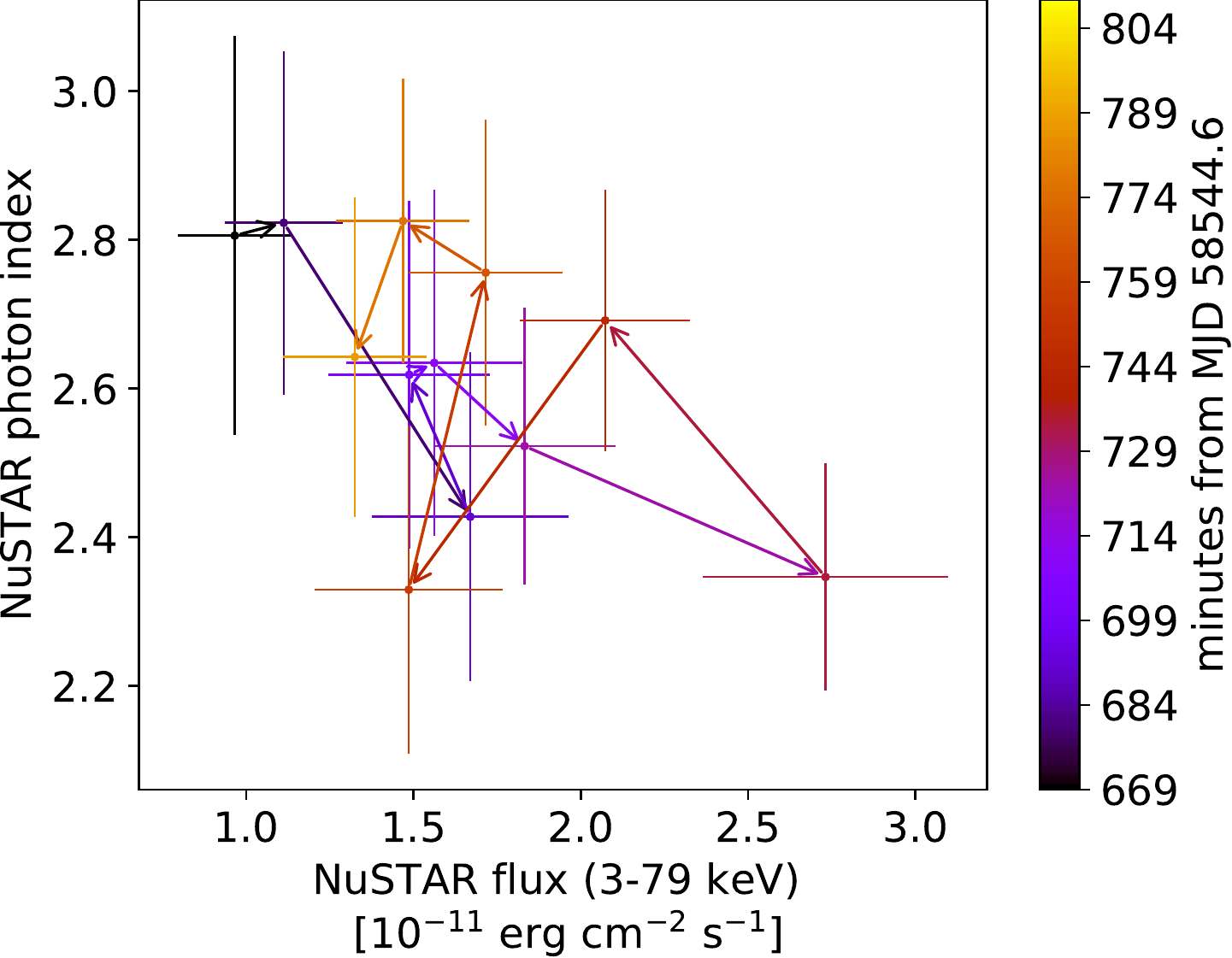}}
\caption{\Xm\ (top) and \Nu\ (bottom) photon index versus integral flux for \source\ during short X-ray flares. The coloured arrows represent the time evolution.} 
\label{fig:hysteresis}
\end{figure}

We then focused our search on the most prominent flares in the X-ray light curves (i.e. the one we also used to constrain the variability time scale), see Fig. \ref{fig:hysteresis}. We slightly narrowed the time window with respect to the time range shown in red in Figs. \ref{fig:exp_fit_xmm} and \ref{fig:exp_fit_nu} focusing  on the most prominent flare time slots, MJD 58544.00 - 58544.08, (i.e. 25 - 125 minutes after the beginning of the observations) for \Xm\ and MJD 58545.08 - 58545.17, (i.e. 670 - 810 minutes after the beginning of the observations) for \Nu. In this time epoch, the data are statistically consistent with a simple harder-when-brighter trend. Indeed, we fitted the data in both epochs with a linear function and we found that the reduced $\chi^2$ is 0.97 for the \Xm\ flare and 0.55 for the \Nu\ flare. 

\subsection{Constraints on emitting region parameters from variability timescales}
\label{constrain_r_and_b}
Using the shortest time scale from these variability studies for individual epochs, the size of the emission region can be constrained  as a function of the Doppler factor $\delta$:
\begin{equation}
\label{size}
R \leq \frac{ct_\textrm{var}\delta}{(1 + z)},
\end{equation}
where we used $t_\textrm{var} = 159 \pm 29$ min for the \Xm\ epoch and $t_\textrm{var} = 86 \pm 18$ min for the \Nu\ epoch (see Table \ref{tab:burstfit}).
    
Since we have no strong constraints on the Doppler factor neither from our observing campaign nor from the previous observations of the source, we used $\delta\sim20$ first, but then iterated the value during the SED modelling (see Section \ref{sect:sedmodeling}). The values close to 20 are rather typical for VHE gamma-ray emitting BL Lacs \citep{tavecchio_2010}.
    
As we were able to constrain the variability timescales in both epochs in the two different energy bands, we followed \cite{Zhang:2002ff} to use these timescales to constrain the magnetic field strength of the emission region. The synchrotron cooling time of electrons is inversely proportional to the square root of the energy of the photons. Indeed, the results of our fit are consistent with this scenario, as we find the \Xm\, lower energy photons cooling time is longer than the one found for the \Nu\, higher energy ones.
    
The magnetic field strength is then calculated using the formula from \cite{Zhang:2002ff}:
\begin{equation}
\label{magn_field}
B = 210 \times \left( \frac{1 + z}{E_l\times\delta}\right)^{1/3} \left[ \frac{1 - \left(E_l / E_h\right)^{1/2}}{\tau_\textrm{soft}}\right]^{2/3}\quad \textrm{G}\,
\end{equation}

where $E_l$ and $E_h$ are taken as the logarithmic mean energies in the low and high-energy band in units of keV for the low-energy band and the high-energy band considered, and $\tau_\textrm{soft}$ is the difference in the decay time values for the energy bands considered. We decided to combine the observations from the two epochs because the time lag in our observations is not statistically significant. By working under the assumption that the magnetic field does not vary between the two epochs, we used the low-energy band of \Xm\ and the high-energy band of \Nu, and $\tau_\textrm{soft} = t_\textrm{decay,XMM} - t_\textrm{decay,NuSTAR}  = 10.56\pm 2.20\, \textrm{ks}$.
    
The upper limit value of $R$ and the estimated value of $B$ corresponding to $\delta = 20$ obtained with the variability timescales estimated from the two epochs are reported in Table \ref{tab:RandB}. These values were used to model the SED, as we will describe in Section \ref{sect:sedmodeling}.

\begin{table}
\caption{Results of the variability time-scale fit and estimation of $R$ and $B$.} 
\label{tab:RandB}
\begin{center}
\begin{tabular}{cccc}
\hline
\multicolumn{1}{c}{Epoch} &
\multicolumn{1}{c}{$\delta$} &
\multicolumn{1}{c}{$R$} &
\multicolumn{1}{c}{$B$}\\
& & $(\times 10^{15} \textrm{cm})$ & (G)
\\
\hline
\emph{XMM} & 20 & 5.07 $ \pm$ 0.92 & \multirow{2}{*}{0.14 $ \pm$ 0.02}\\
\Nu\ & 20 & 2.73 $\pm$ 0.58 & \\
\hline
\end{tabular}
\end{center}
\end{table}
    
\section{Spectral energy distribution}
\label{spectralenergydistribution_section}
In this campaign, we measured the SED of \source\ for the first time from the optical to the VHE gamma-ray band. This allowed us to investigate the source classification and possible emission models during the flaring state.

\subsection{Source classification}
\label{classification}

As reported in \cite{Biteau:2020prb}, it is possible to distinguish three kind of extreme behaviours: HBL sources showing extreme behaviour during flaring states, when synchrotron and IC peaks shift towards higher frequencies, going back to their standard HBL-like state, as observed for Mrk501 \citep{GHISELLINI1999}; sources with a steady hard synchrotron without evidence for a hard spectrum at TeV energies; finally, sources showing a persistently hard IC hump peaking at and above TeV energies and a synchrotron peak in the X-ray band. Our excellent X-ray and VHE gamma-ray data allowed us to search for these signatures in the SEDs that we measured at different times for \source.

\source\ was very little studied before the flare occurred in 2019. As explained in the previous Sections, the photon index reported by \cite{4FGL_2019} seemed to suggest a possible EHBL classification. Thus, we examined archival data from the ASI Space Science Data Center (SSDC)\footnote{\url{http://www.asdc.asi.it/}} to test if the source could indeed be classified as such. According to \cite{bonnoli_2015}, a good criterion to select EHBLs relies on the high X-ray--to--radio flux ratio. As reported by the authors, a high ratio of X-ray versus radio flux ($F_X/F_R>10^4$) would indicate a good EHBL candidate. Based on archival data, considering the flux in the $\sim 10^{16} - 10^{19}$ Hz band and in the $\sim 10^9 - 10^{10}$ Hz band for $F_X$ and $F_R$ respectively, we found a ratio of $\sim 30$, which led us to the conclusion that it was not a good EHBL candidate. 

\begin{figure}
 \includegraphics[width=\columnwidth]{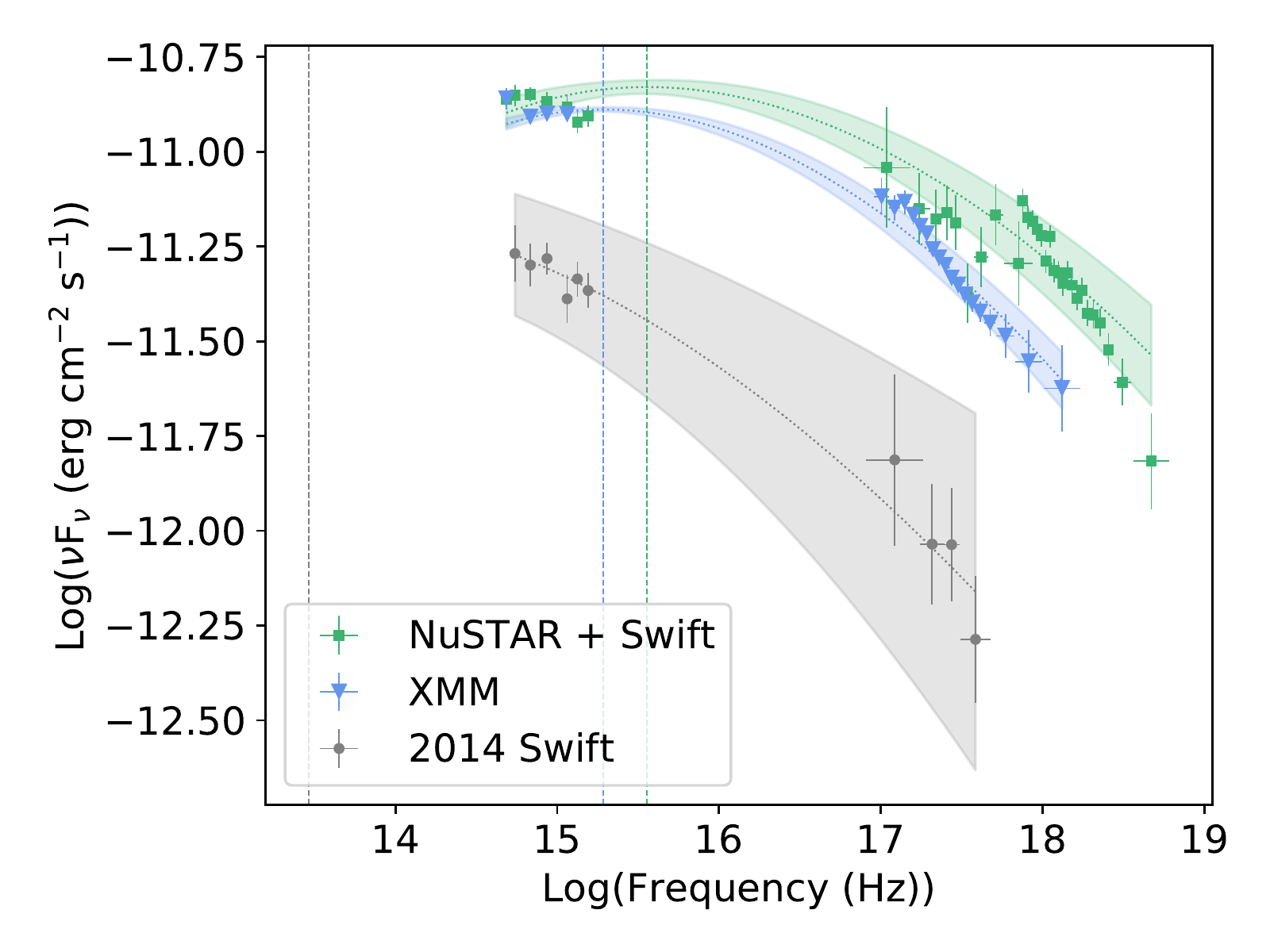}
 \caption{Results of the fit of the SED synchrotron bump with a log-parabola. Data are shown in different colours, representing: \Nu\ epoch (\Nu\ and \emph{Swift} datasets taken on MJD 58544.84 and 58544.97}, green squares), \emph{XMM} epoch (\Xm\ and \emph{XMM-OM} dataset, blue triangles), 
 2014 \emph{Swift} dataset (grey circles). Dotted curves are showed superimposed on SED points with the same colours, representing the fit curves and the uncertainty bands. Vertical dashed line are also shown, in correspondence of the peak frequency evaluated from the fit parameters.
 \label{fig:peak_fit}
\end{figure}

In order to estimate the peak frequency and classify the source, we combined strictly simultaneous data in the optical and X-ray band to perform a fit of a log-parabola to the data in the log-log plane:
\begin{equation}
    \nu F_\nu(\nu) = f_0 \cdot 10^{- b \cdot (\log_{10}(\nu/\nu_\textrm{s}))^2},
\end{equation}
where $\nu_\textrm{s}$ is the peak frequency. 
This fit was performed separately on data from the current observations and a previous \emph{Swift}-UVOT and \emph{Swift}-XRT observations from 2014. We considered data from the two epochs separately in order to take into account a possible shift in the peak frequency between the two observing periods, combining \emph{Swift}-XRT observations from MJD 58544.84 and 58544.97 with the \Nu\ dataset.
The results of the fit procedure on the three considered datasets are reported in Fig. \ref{fig:peak_fit}. The best fit values for the three datasets are reported in Table \ref{pol3_table}.

\begin{table}
\caption{Results of the log-parabola fit of the synchrotron peak of the SED of the three different epochs considered.} 
\label{pol3_table}
\begin{center}
\begin{tabular}{cc}
\hline
\multicolumn{1}{c}{Epoch} &
\multicolumn{1}{c}{Log of peak frequency (Hz)} \\
\hline
2014 \emph{Swift}  & $13.46\, \pm \,2.53$\\
\emph{XMM} & $15.28\, \pm0.06$\\
\Nu\ and 2019 \emph{Swift} & $15.56\, \pm \, 0.11$\\
\hline
\end{tabular}
\end{center}
\end{table}

The best fit values found for the most recent observations classify the source as HSP during the flare. For 2014 data, however, the value found $\nu_\textrm{s} \simeq 10^{13.46\, \pm \,2.53} $ Hz does not allow any conclusion to be drawn on the classification of the source.
However, it is clear that  $\nu_\textrm{s}$ moved to higher energy between the \Xm\ and the \Nu\ observations, which were separated by less than a day. This is most naturally explained by the injection of fresh electrons or the dominance of a new emission component. Even in the flaring state detected by \Nu\ the peak frequency does not exceed $10^{17}$ and therefore it is clear that \source\ is not an extreme synchrotron peaked source.

\subsection{Emission models}
\label{emissionmodels}
Most blazars' observations are fitted with radiative models. These models are usually classified as leptonic and hadronic models.

In the simple one-zone SSC model, TeV emission is the result of the IC scattering of electrons and positrons in the jet on photons created by the electron population itself via synchrotron emission. The SSC model is supported by several observations of good temporal correlation between the TeV and X-ray flares \citep[see][]{Maraschi_1999,Takahashi_2000, Krawczynski:2001tw,Coppi_1999}.
In this simple one-zone SSC model the energy density has been found to be largely dominated by particle energy density rather than the magnetic field; this unusually low magnetisation seems to be in contradiction with theoretical and observational constraints of equipartition conditions, which cannot be reproduced in BL Lac objects with the one-zone model \citep{2016MNRAS.456.2374T}. It was recently suggested, however, that equipartition can be achieved in one-zone models via the introduction of an anisotropic electron population \citep{2020MNRAS.491.2198T}. 
The one-zone SSC is still the default model and, given the fact that during this campaign all of the energy bands were found to be in the same high state, we apply this model to the data using the constraints for the size of the emission region and magnetic field strength we found in Section \ref{constrain_r_and_b}.

Another possible way to solve the contradiction on the low magnetization can be obtained by taking into consideration the  existence of additional seed photons from other parts of the jet \citep[e.g.][]{Georganopoulos_2003,ghisellini2005}. In these models one assumes the jet to be structured.
As shown in \cite{tavecchio_ghisellini2015}, the assumption of a supplementary source of soft photons intervening in the IC emission allows the reproduction of the observed SED assuming equipartition between the magnetic and the electron energy densities. 
There is also observational evidence for spine-sheath models from very long baseline interferometry observations \citep[e.g.][]{attridge99, giroletti04} and indications for multiple components contributing to the optical band from long-term variability \citep{lindfors16} and polarization \citep[see e.g.][]{valtaoja91}.

In our case, the X-ray variability indicates that the X-ray emission region is very compact. Therefore, it must either be located very close to the central engine (to fill the full diameter of the conical jet) or be embedded in a larger emission region. Our X-ray light curves show several flares and the SED peak moves to higher energies from the \Xm\ epoch to the \Nu\ epoch. This can be either due to the variation of the particle distributions within the emission region (e.g. fresh injection of particles) or due to the flaring region (filling the full diameter of the jet) consisting of several emission regions. 
Therefore, we also model the SEDs with a two-component model such that we keep the parameters of the larger region constant between the two SED epochs and only vary the parameters of the small emission region.

Another possible explanation of the VHE emission observed is given by hadronic models \citep[e.g.][]{1993A&A...269...67M, 2001APh....15..121M, 2010APh....34..258A, cerruti2015}, where energetic photons are produced in jets via hadronic interactions. These models  are based on the assumption of the presence of high-energy protons in the jet which are accelerated together with the electrons. While the low-energy peak is still explained by synchrotron radiation of electrons, in this scenario the VHE gamma rays are thought to be produced by interactions of the relativistic protons with soft photons or with the magnetic field. 
In particular, the recent indication of a link between a neutrino track event and the blazar TXS\,0506+056 \citep{IceCube:2018dnn,IceCube:2018cha} can be interpreted in a lepto-hadronic scenario where electrons and protons are accelerated in the jet, and synchrotron photons from the electrons lead to photo-hadronic neutrino production \citep{Keivani_2018, 2018ApJ...863L..10A, Cerruti:2018tmc}. 

However, since our study shows that \source\ is a rather typical HSP in flaring state and since no neutrinos have been detected from the direction of \source\ \citep{2020PhRvL.124e1103A}, in the next Section we investigate only one-zone and two-component leptonic models for the spectral energy distribution.

\subsection{Spectral energy distribution modelling}
\label{sect:sedmodeling}
\begin{figure*}
\centering
\subfloat[][]
 {\includegraphics[width=\columnwidth]{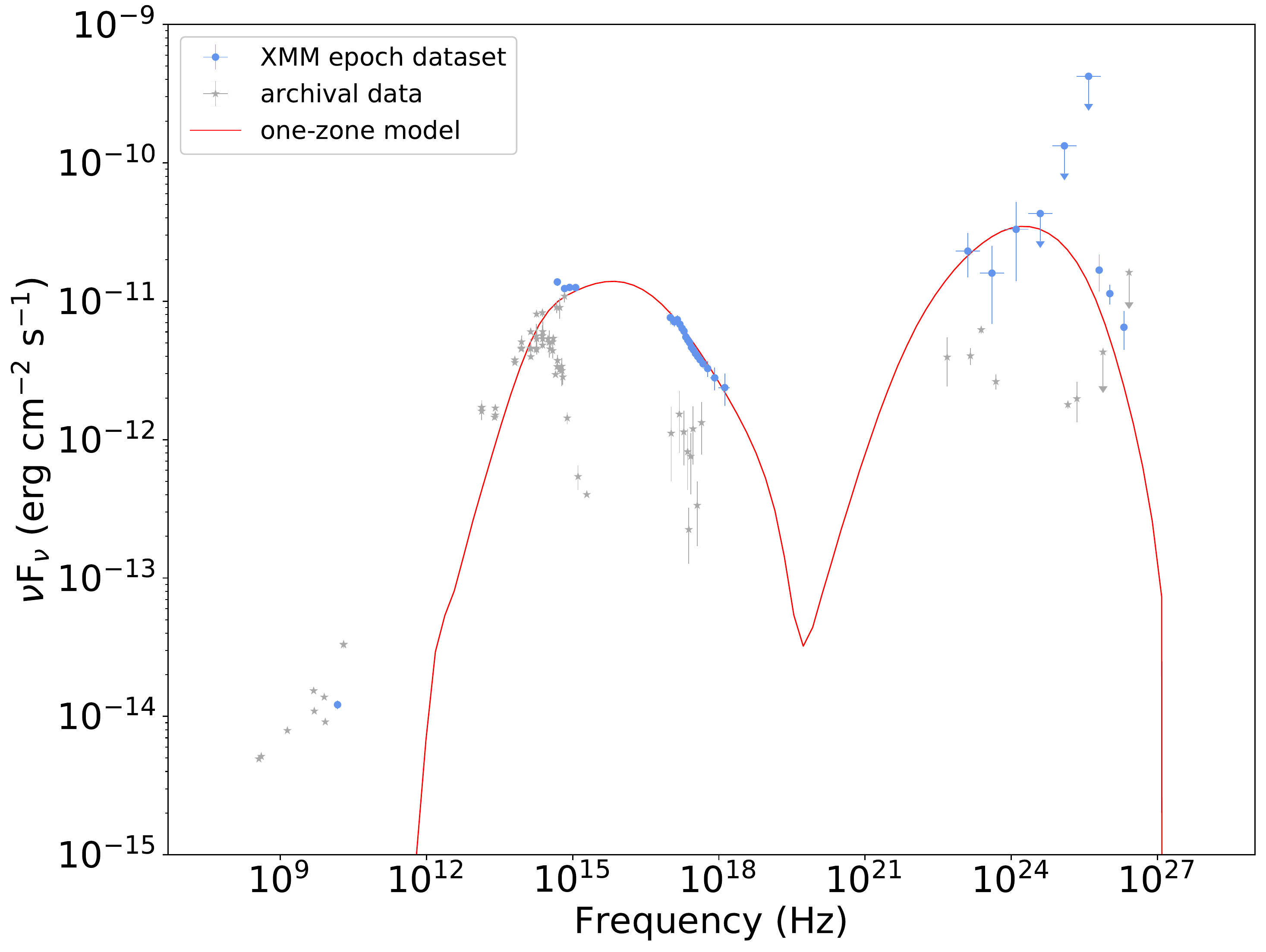}\label{fig:xmm_one}}\qquad
 \subfloat[][]
 {\includegraphics[width=\columnwidth]{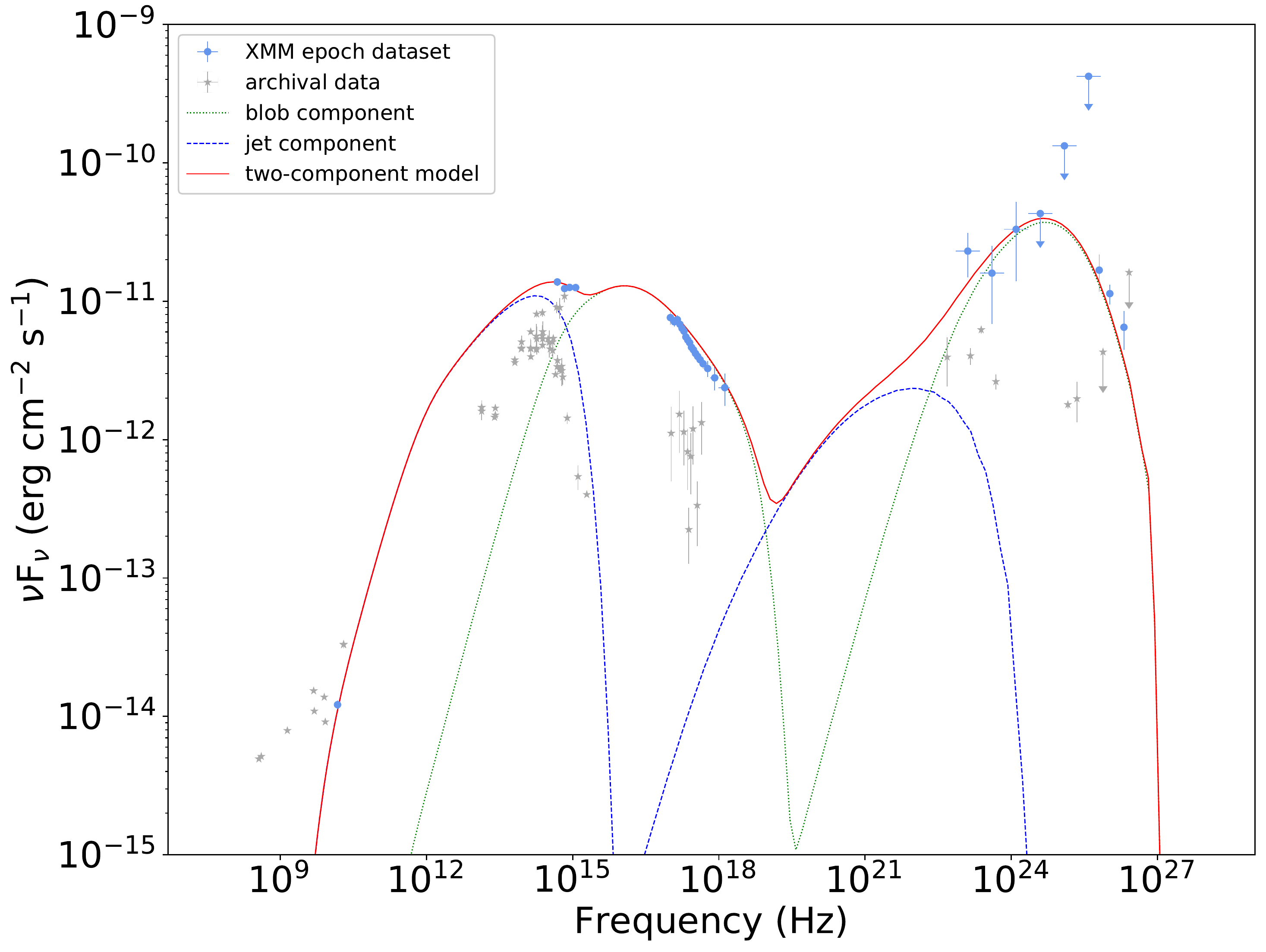}\label{fig:xmm_two}}
 \caption{Broadband SED and modelled spectra for \source. The SED has been modelled considering the VLBA SED point in the radio band, KVA,  \emph{XMM-OM} and \Xm\ SED points in the optical and X-ray energy band and the \emph{Fermi}-LAT and MAGIC SED points in the gamma-ray band}, using a one-zone model (left) and a two-component model (right). Coloured points represent observations in the different energy bands during the observations of 2019, grey points represent archival data. The red solid line represents the models, the green dotted line represents the blob emission and the blue dashed line represents the jet emission. Further details can be found in the text. See Table \ref{tab:sedmodel} for parameters.
 \label{fig:xmm_modeling}
\end{figure*}

\begin{figure*}
\centering
\subfloat[][]
 {\includegraphics[width=\columnwidth]{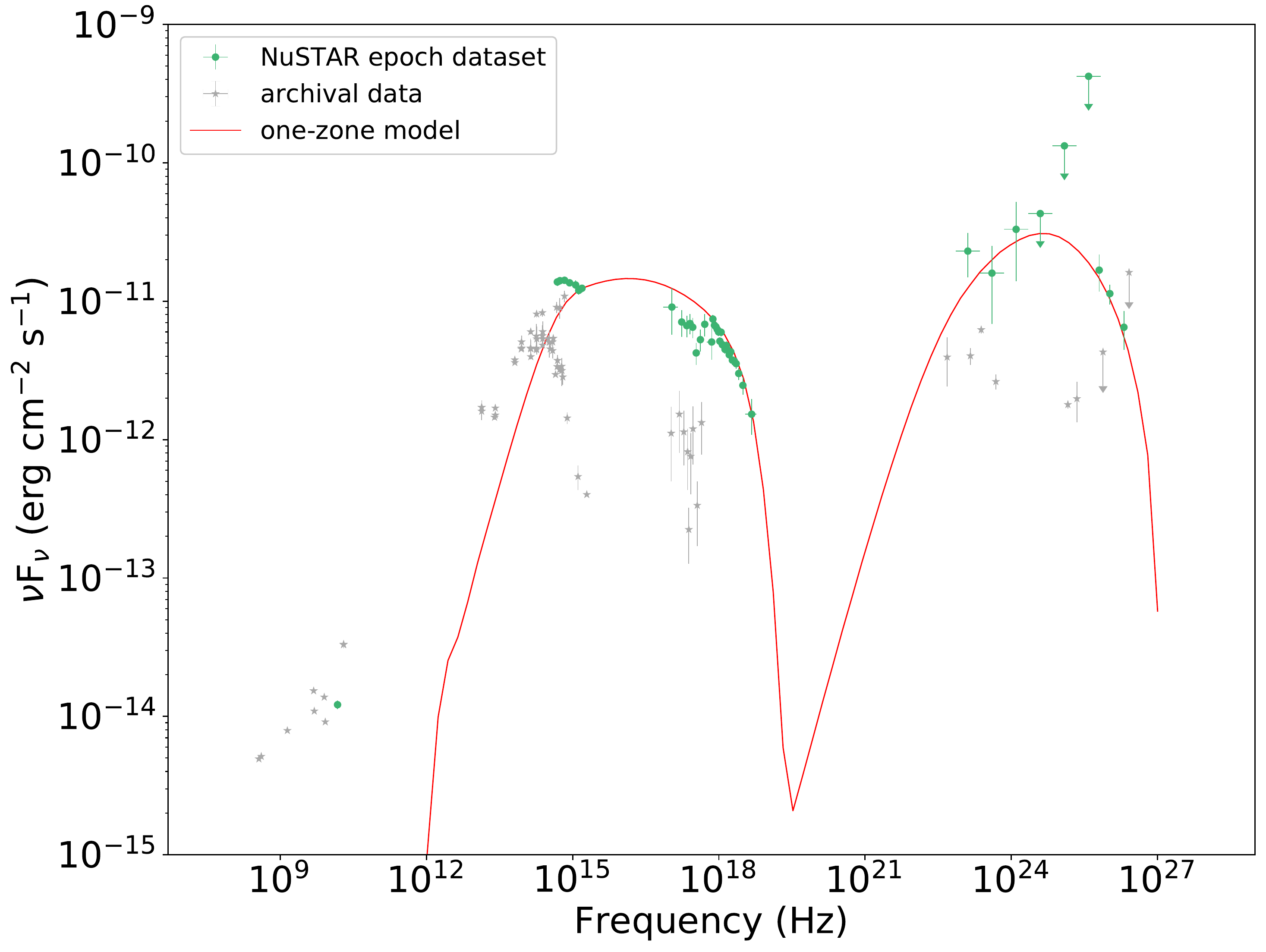}\label{fig:nu_one}}\qquad
 \subfloat[][]
 {\includegraphics[width=\columnwidth]{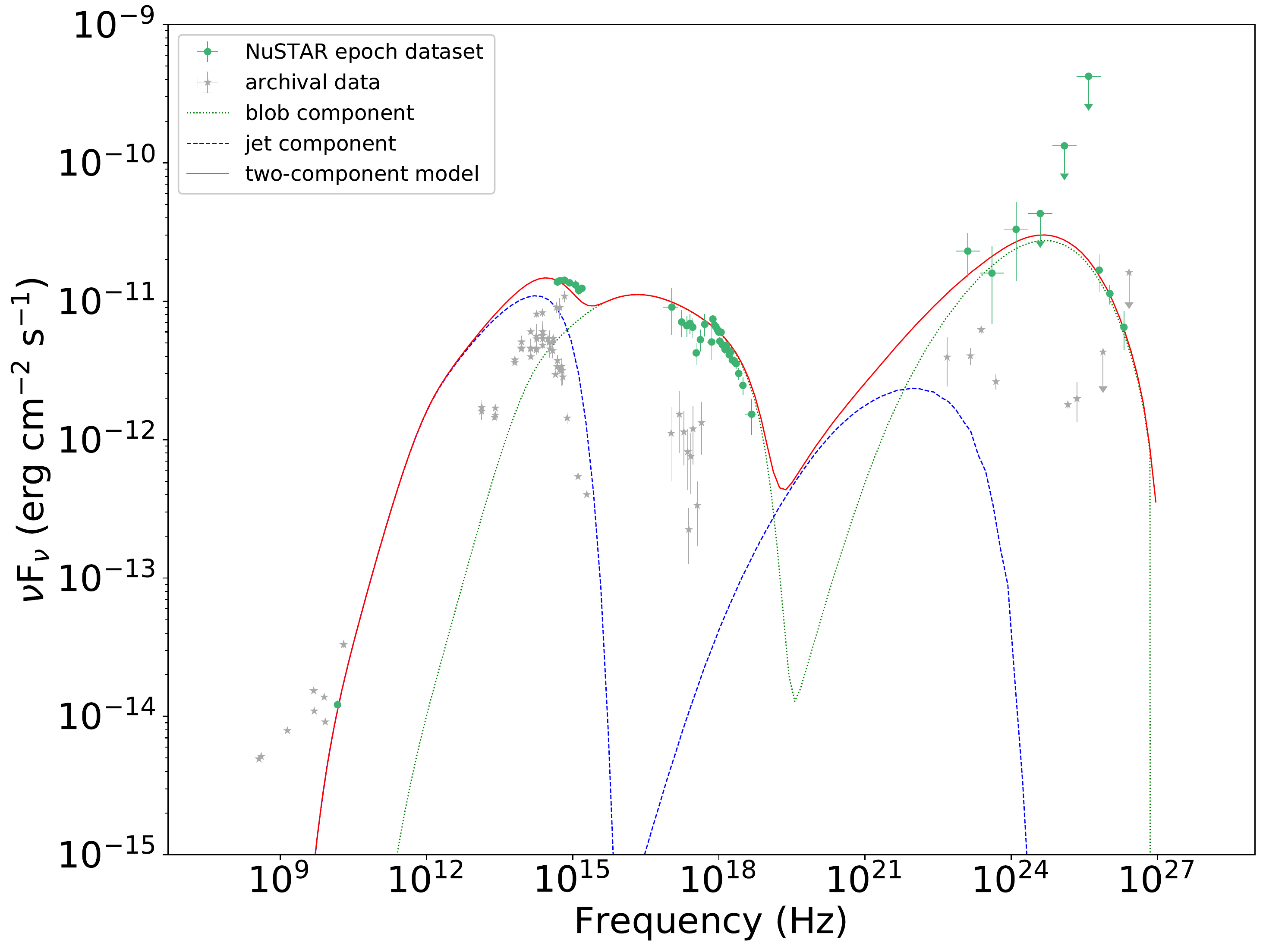}\label{fig:nu_two}}
 \caption{Broadband SED and modelled spectra for \source. The SED has been modelled considering the VLBA SED point in the radio band, KVA, \emph{Swift} and \Nu\ SED points in the optical and X-ray energy band and the \emph{Fermi}-LAT and MAGIC SED points in the gamma-ray band}, using a one-zone model (left) and a two-component model (right). Only \emph{Swift} data simultaneous to \Nu\ observations were used. Coloured points represent observations in the different energy bands during the observations of 2019, grey points represent archival data. The red solid line represents the models, the green dotted line represents the blob emission and the blue dashed line represents the jet emission. Further details can be found in the text. For parameters, see Table \ref{tab:sedmodel}.
 \label{fig:nustar_modeling}
\end{figure*}

We derived the broadband SED from radio to VHE shown in Figs \ref{fig:xmm_modeling} and \ref{fig:nustar_modeling}. For comparison purposes, archival data from the SSDC and from a previous \emph{Swift}-XRT detection are also shown as grey dots.

The light curve in the \emph{Fermi}-LAT band showed the presence of a high flux state on MJD 58546. Since no VHE observations were available on that day, we performed the \emph{Fermi}-LAT spectral analysis starting from MJD 58541 up to MJD 58545 in order to have a smooth connection between the HE and the VHE gamma-ray observations. 
For what concerns the optical and X-ray observations, since these were not simultaneous and the source showed short-term variability in the X-ray energy range, we decided to  separate the datasets into \Nu, \emph{Swift}-XRT and \emph{Swift}-UVOT observations and \Xm\ and \emph{XMM-OM} observations. We modelled these datasets separately since the model we adopted is not time-dependent.

Given the fact that the source was in a high state in all energy bands, the  SED was at first modelled with a simple one-zone SSC model; the radiation is emitted in a region in the jet by a single homogeneous population of electrons \citep{Maraschi_2003} which is responsible for the emission from infrared to VHE frequencies. 
The emitting region can be described as a sphere with radius $R$ with a uniform magnetic field $B$. The Doppler factor, $\delta$, is required to take into account the relativistic effects. For HSP sources, this factor is usually $\sim 10 - 50$: within this range we selected a value for each epoch that would reproduce a model in good agreement with our data, having a consistent model by using the same value of $\delta$ to estimate $R$ and $B$.

A good agreement between the model and the SED data was found using $\delta=18$ for the \Xm\ epoch, which corresponds to an estimated magnetic field intensity $B = 0.14 \pm 0.02\,\textrm{G}$ and to an upper-limit on the size equal to $R \leq 4.56 \pm 0.83 \times 10^{15}\, \textrm{cm}$, and $\delta = 24$ for the \Nu\ epoch, which corresponds to $B = 0.13 \pm 0.02\,\textrm{G}$ and $R \leq 3.28 \pm 0.69 \times 10^{15}\, \textrm{cm}$.

The population of relativistic electrons is described by a broken power-law model, where $K$ is a normalisation factor which represents the density of electrons with $\gamma = 1$, $n_1$ the index from $\gamma_\textrm{min}$ to $\gamma_\textrm{br}$ and $n_2$ from $\gamma_\textrm{br}$ to $\gamma_\textrm{max}$, for a total of six parameters. The model includes the Klein-Nishina cross section $\sigma_{KN}$ for the IC spectrum calculation, which is important in the case of emission above the GeV range. Since our analyses in Sections \ref{spectral_ev} and \ref{classification} clearly demonstrated that between the \Xm\ and \Nu\ observing epochs the synchrotron peak moved to higher energies and spectral evolution with time changed the pattern, we assume that the new blob dominates the emission in the \Nu\ epoch, i.e. we let the electron energy spectrum parameters vary rather arbitrarily between the two epochs. 

For both datasets, we found acceptable agreement between the models and the data available from optical to VHE gamma-ray (see Figs. \ref{fig:xmm_one} and \ref{fig:nu_one}). The parameters of the one-zone model are given in Table \ref{tab:sedmodel}.
Even if we tried a large number of different combinations of parameters, it was hard to find parameters that would reproduce well the shape of the synchrotron bump. Parameters that would describe well the shape of the spectrum in the X-ray band led to an underestimation of the optical flux while changing the set of parameters so that the optical shape was well described by the model would overproduce the X-ray flux. Moreover, we also had to take into account the IC bump which extends to VHE gamma-ray energies.
In the \Xm\ epoch, the biggest challenge for the model is to reproduce the rather soft \Xm\ spectra and still let the second peak extend to VHE gamma-rays without overproducing the HE gamma-ray part. The selected model is a compromise and is below the VHE gamma-ray points. 
In the \Nu\ epoch, the selected model produces well the level of the flux in the UV-band and the shape and level of the \Nu\ observations, but does not describe the shape in the optical band and overproduces the \emph{Swift}-XRT data. The second bump is well described in the \Nu\ epoch.

As expected, the one-zone model is not able to reproduce the emission at radio frequencies (15 GHz): this is due to the small and dense emission region. The emission foreseen by the one-zone model is synchrotron self-absorbed. It is likely that the radio flux originates from a different, larger, region of the jet, transparent at those frequencies. Also, as is typical for one-zone models, both the \Xm\, and \Nu\, one-zone models are far from equipartition (see Table \ref{tab:sedmodel}, last column).

The two-component model could potentially improve the modelling of the synchrotron component and there exists numerous observational evidence supporting two-component models for BL~Lac objects in general (see Section~\ref{emissionmodels}). Therefore we modelled the observed SEDs with the two-component model described in \cite{2020arXiv200604493M}, and references therein. The model includes two spherical blobs embedded in one another which are interacting. Each of the blobs is filled with relativistic electrons with a broken power-law distribution. The two blobs have the same magnetic field strength, but different Doppler factors and sizes, increasing the number of parameters from eight to fifteen. We call the two emission regions "blob" (smaller region) and "jet" (larger region). The interaction between the emission regions provides additional seed photons for Compton scattering. The larger emission region dominantly provides seed photons to the smaller region \citep{2011A&A...534A..86T}. The gamma-gamma absorption is negligible \citep[see][]{2020arXiv200604493M}. The two-component model describes the data from the radio to the VHE range well (see Figs. \ref{fig:xmm_two} and \ref{fig:nu_two}) with the parameters given in Table \ref{tab:sedmodel}. The improvement with respect to the one-zone model is most evident in the radio and optical part. While the lowest radio frequencies are still not reproduced in the two-component model, the "jet" component connects smoothly the 15\,GHz MOJAVE point to the optical data and the shape of the optical SED is reproduced better. The jet component is closer to equipartition, but we do not find solutions where the blob component itself would be in equipartition; indeed, for \Xm\, and \Nu\, we find values of the order $10^{-3}-10^{-4}$. The main difficulty is to produce the Compton dominance we see in the observed SED.

\begin{table*}
\caption{SED modelling parameters for one-zone SSC and two-component models. Parameters are reported for the two available X-ray data set. See the text for the description of the parameters and the models.}
\label{tab:sedmodel}
\resizebox{\textwidth}{!}{
\begin{tabular}{ccccccccccccc}
\hline
X-ray & Model & Component & $\gamma_\textrm{min}$ & $\gamma_\textrm{b}$ & $\gamma_\textrm{max}$ & $n_1$ & $n_2$ & $B$ & $K$ & $R$ & $\delta$ & $U_\textrm{B}/U_\textrm{E}$\\
dataset & & &	$(\times 10^3)$ & $(\times 10^4)$ & $(\times 10^5)$ & & & (G) & $(\times 10^{4} \textrm{cm}^{-3})$ & $(\times 10^{15} \textrm{cm})$ & &\\
\hline
\multirow{3}{*}{\Xm} & one-zone & - & 4.3 & 3.1 &  10.0 & 2.21 &  3.95 &  0.14 &  59 &  4.56 &  18 & 5 $\times 10^{-3}$\\
& \multirow{2}{*}{2-component}    & blob & 6.0 & 3.5 & 6.0 & 1.96 & 4.0 & 0.14 & 6.1 & 3.95 & 18 & 4 $\times 10^{-3}$\\
& & jet & 0.52 & 0.52 & 0.19 & 1.74 & 2.84 & 0.14 & 0.0041 & 320 & 3.8 & 1.4\\
\hline
\multirow{3}{*}{\Nu} & one-zone & - & 4.9 & 5 &  4.5 & 2.51 &  3.72 &  0.14 &  980 &  3.22 &  24 & 6 $\times 10^{-3}$\\
& \multirow{2}{*}{2-component} & blob & 2.5 & 2.7 & 5.8 & 1.99 & 3.5 & 0.14 & 9.3 & 2.92 & 22 & 3 $\times 10^{-3}$\\
& & jet & 0.52 & 0.52 & 0.19 & 1.74 & 2.84 & 0.14 & 0.0041 & 320 & 3.8 & 1.4\\
\hline
\end{tabular}}
\end{table*}

\section{Summary and Discussion}
\label{summary_sect}
In this paper we present a MWL analysis of the source \source, a BL Lac object which had been very little studied before this campaign. When compared to archival data in the X-ray and HE gamma-ray energy bands and optical monitoring following the analysed period, at the end of February 2019 the source was found to be in a flaring state in all energy bands, which lead to its first detection in the VHE gamma-ray band by MAGIC. Since this time, no further TeV observations have been published so no additional information on its VHE gamma-ray emission is available.

Thanks to the very good X-ray coverage of the source during the flaring state it was possible to detect an hour-scale variability in the X-ray fluxes and a clear change in the spectral shape between the \Xm\ (MJD 58544) and \Nu\ (MJD 58545) observations. We produced the \Xm\ and \Nu\ light curves in two energy bins each and constrained the rise and decay times of the flares. We find rapid variations on the time scale of the order of hours in both datasets.
Studies of variability of blazars in the \Nu\ band \citep{2018Bhatta} show that the timescale we found is in line with the fastest variability seen in many HSP sources. 
In few cases, even shorter timescales variability has been found, e.g., for Mrk 421  \citep{2020arXiv200108678M} and for 1ES\,1959+650 \citep{2020A&A...638A..14M}, although different analytical functions were adopted in these works to profile the flares. The short time-scale variability was used to constrain the size of the emission region and the intensity of the magnetic field, both used later during the analysis to find a suitable model for the broadband emission.

\source\ was very little studied prior to this work and therefore the source had not yet been properly classified. Up to and including the 3FGL \emph{Fermi}-LAT catalog, the source was still classified as a blazar candidate of uncertain type. Only in the 4FGL catalog was it classified as a BL Lac object, but the SED type was not defined. The good coverage in the optical--X-ray energy band obtained in this observational campaign allowed an accurate determination of the synchrotron peak frequency $\nu_\textrm{s}$ and the classification of the source as a HSP source during this flare. Furthermore, during the flare, the peak clearly shifted to higher energies in a timescale of less than a day. Such behaviour is rather common in blazars, see e.g. \cite{1998ApJ...492L..17P} or \cite{2020A&A...638A..14M}, in which a clear shift in the synchrotron bump was reported during flare activity with respect to archival data. In our case, thanks to very good X-ray coverage  we could follow this shift on daily time scales.

We also investigated the behaviour of the jet at 15\,GHz using the VLBA data from the MOJAVE program. We found no new components or moving components, which is in line with what is seen in other TeV blazars with high synchrotron peak frequency \citep[see e.g.][and references there in]{piner18}. HSP objects usually do not show high linear core polarization levels \citep{lister11}, which is what we also find for TXS~1515-273 in the first epochs of data. However, in 2019 the core was significantly polarized, which might be connected to the general high state, but the sparse coverage of the VLBA observations did not allow a strong conclusion to be drawn.

We also modelled the broadband SED from radio to VHE gamma rays. The evolution of the SED from the \Xm\ to the \Nu\ epoch clearly suggests that an injection of new electrons or a new blob is starting to dominate the emission. The one-zone model describes well the data from the X-ray band to VHE gamma rays for both epochs, but has problems reproducing the shape of the optical part of the SED and does not reproduce the radio. The latter is expected as the small emitting region considered is optically thick to radio emission. The radio emission must then originate from a different component. These issues are solved when a two-component model is considered, in which the blob is responsible for the emission from X-ray to VHE gamma rays, while the jet dominates the emission at radio wavelengths. The introduction of the two-component model increases the number of model parameters from eight to fifteen, allowing for a more accurate modelling of the emission. In particular, the jet emission results in important contributions in the optical band. Moreover, assuming the two emission regions to be co-spatial, seed photons for the inverse Compton scattering are provided also from the jet.

Finally, we examined the ratio between the magnetic field density and the electron energy density, $U_\textrm{B}/U_\textrm{E}$. In both epochs, we find that the one-zone SED parameters are quite far from equipartition, with the magnetic field energy density dominated by the kinetic energy of the relativistic particles by several orders of magnitude. In the two-component model instead the jet itself is in equipartition, but within the constraints on emission region size and magnetic field strength we derived from the X-ray observations (assuming "typical" Doppler factors), we did not find solutions where the blob would be in equipartition. We note that \cite{2016MNRAS.456.2374T} found a two-component model solution for the low state SED of Mrk421 where the blob itself was in equipartition. However, this seems not to always be possible in the case of flares \citep[see also][]{bllac}.

\section*{Acknowledgements}
We would like to thank the Instituto de Astrof\'{\i}sica de Canarias for the excellent working conditions at the Observatorio del Roque de los Muchachos in La Palma. The financial support of the German BMBF, MPG and HGF; the Italian INFN and INAF; the Swiss National Fund SNF; the ERDF under the Spanish Ministerio de Ciencia e Innovaci\'{o}n (MICINN) (FPA2017-87859-P, FPA2017-85668-P, FPA2017-82729-C6-5-R, FPA2017-90566-REDC, PID2019-104114RB-C31, PID2019-104114RB-C32, PID2019-105510GB-C31,PID2019-107847RB-C41, PID2019-107847RB-C42, PID2019-107847RB-C44, PID2019-107988GB-C22); the Indian Department of Atomic Energy; the Japanese ICRR, the University of Tokyo, JSPS, and MEXT; the Bulgarian Ministry of Education and Science, National RI Roadmap Project DO1-268/16.12.2019 and the Academy of Finland grant nr. 317637 and 320045 are gratefully acknowledged. This work was also supported by the Spanish Centro de Excelencia ``Severo Ochoa'' SEV-2016-0588, SEV-2017-0709 and CEX2019-000920-S, and "Mar\'{\i}a de Maeztu” CEX2019-000918-M, the Unidad de Excelencia ``Mar\'{\i}a de Maeztu'' MDM-2015-0509-18-2 and the "la Caixa" Foundation (fellowship LCF/BQ/PI18/11630012), by the Croatian Science Foundation (HrZZ) Project IP-2016-06-9782 and the University of Rijeka Project 13.12.1.3.02, by the DFG Collaborative Research Centers SFB823/C4 and SFB876/C3, the Polish National Research Centre grant UMO-2016/22/M/ST9/00382 and by the Brazilian MCTIC, CNPq and FAPERJ.

The \emph{Fermi}-LAT Collaboration acknowledges generous ongoing support from a number of agencies and institutes that have supported both the development and the operation of the  LAT  as  well  as  scientific  data  analysis.  These  include  the National  Aeronautics  and Space   Administration   and   the   Department   of   Energy   in the  United   States,   the Commissariat à l'Energie Atomique and the Centre National de la Recherche Scientifique/Institut  National  de  Physique  Nucléaire et  de  Physique  des  Particules  in  France,  the Agenzia  Spaziale  Italiana  and  the Istituto  Nazionale  di  Fisica  Nucleare  in  Italy,  the Ministry  of  Education,  Culture, Sports,  Science  and  Technology  (MEXT),  High  Energy Accelerator  Research Organization  (KEK)  and  Japan  Aerospace  Exploration  Agency (JAXA)  in  Japan,and  the  K.  A.  Wallenberg  Foundation,  the  Swedish  Research  Council and the Swedish National Space Board in Sweden. Additional  support  for  science  analysis  during  the  operations  phase  from  the following agencies is also gratefully acknowledged: the Istituto Nazionale di Astrofisica in Italy and and the Centre National d'Etudes Spatiales in France.

This research has made use of data and/or software provided by the High Energy Astrophysics Science Archive Research Center (HEASARC), which is a service of the Astrophysics Science Division at NASA/GSFC.
This research has made use of data from the MOJAVE database that is maintained by the MOJAVE team \citep{Lister_2018}.

\section*{DATA AVAILABILITY}

The data underlying this article will be shared on reasonable request to the corresponding author.



\bibliographystyle{apalike}

\section*{Affiliations}

(1) Instituto de Astrofisica de Canarias, La Laguna (Tenerife), Spain, (2) Universit\`a di Udine and INFN Trieste, Italy, Udine, Italy, (3) INAF - National Institute for Astrophysics, Roma, Italy, (4) ETH Z\"urich, Institute for Particle Physics, Z\"urich, Switzerland, (5) Institut de F\'isica d'Altes Energies (IFAE), The Barcelona Institute of Science and Technology (BIST), E-08193 Bellaterra (Barcelona), Spain, Bellaterra (Barcelona), Spain, (6) Japanese MAGIC Group: ICRR, The University of Tokyo, Chiba, Japan, (7) Technische Universit\"at Dortmund, Dortmund, Germany, (8) Croatian MAGIC Group: University of Zagreb, Faculty of Electrical Engineering and Computing (FER), Zagreb, Croatia, (9) IPARCOS Institute and EMFTEL Department, Universidad Complutense de Madrid, Madrid, Spain, (10) Centro Brasileiro de Pesquisas F\'isicas (CBPF), Rio de Janeiro, Brazil, (11) Dipartimento di Fisica e Astronomia dell?Universit\`a and Sezione INFN, Padova, Italy, Padova, Italy, (12) University of Lodz, Faculty of Physics and Applied Informatics, Department of Astrophysics, 90-236 Lodz, Poland, Lodz, Poland, (13) Dipartimento SFTA, Sezione di Fisica, Universit\`a di Siena and INFN sez. di Pisa, Siena, Italy, (14) Deutsches Elektronen-Synchrotron (DESY) Zeuthen, Zeuthen, Germany, (15) INFN MAGIC Group: INFN Sezione di Torino and Universit\`a degli Studi di Torino, Torino, Italy, (16) Max-Planck-Institut f\"ur Physik, M\"unchen, M\"unchen, Germany, (17) Universita di Pisa and INFN Pisa, Pisa, Italy, (18) Universitat de Barcelona, Barcelona, Spain, (19) Armenian MAGIC Group: A. Alikhanyan National Science Laboratory, Yerevan, Armenia, (20) Centro de Investigaciones Energeticas, Medioambientales y Tecnologicas, Madrid, Spain, (21) INFN MAGIC Group: INFN Sezione di Bari and Dipartimento Interateneo di Fisica dell'Universit\`a e del Politecnico di Bari, Bari, Italy, (22) Croatian MAGIC Group: University of Rijeka, Department of Physics, Rijeka, Croatia, (23) Institut f\"ur Theoretische Physik und Astrophysik, Fakult\"at f\"ur Physik und Astronomie, Universit\"at W\"urzburg, W\"urzburg, Germany, (24) Finnish MAGIC Group: Finnish Centre for Astronomy with ESO, University of Turku, Turku, Finland, (25) Universitat Aut\`onoma de Barcelona, Barcelona, Spain, (26) Armenian MAGIC Group: ICRANet-Armenia at NAS RA, Yerevan, Armenia, (27) Croatian MAGIC Group: University of Split, Faculty of Electrical Engineering, Mechanical Engineering and Naval Architecture (FESB), Split, Croatia, (28) Croatian MAGIC Group: Josip Juraj Strossmayer University of Osijek, Department of Physics, Osijek, Croatia, (29) Japanese MAGIC Group: RIKEN, Saitama, Japan, (30) Japanese MAGIC Group: Department of Physics, Kyoto University, Kyoto, Japan, (31) Japanese MAGIC Group: Department of Physics, Tokai University, Kanagawa, Japan, (32) Saha Institute of Nuclear Physics, HBNI, Kolkata, India, (33) Institute for Nuclear Research and Nuclear Energy, Sofia, Bulgaria, (34) Finnish MAGIC Group: Astronomy Research Unit, University of Oulu, Oulu, Finland, (35) Croatian MAGIC Group: Ru?er Bo?kovi? Institute, Zagreb, Croatia, (36) INFN MAGIC Group: INFN Sezione di Perugia, Perugia, Italy, (37) INFN MAGIC Group: INFN Roma Tor Vergata, Roma, Italy, (38) now at University of Innsbruck, (39) also at Port d'Informaci\'o Cient\'ifica (PIC) E-08193 Bellaterra (Barcelona) Spain, (40) now at Ruhr-Universit\"at Bochum, Fakult\"at f\"ur Physik und Astronomie, Astronomisches Institut (AIRUB), 44801 Bochum, Germany, (41) also at Dipartimento di Fisica, Universit\`a di Trieste, I-34127 Trieste, Italy, (42) Max-Planck-Institut f\"ur Physik, D-80805 M\"unchen, Germany, (43) also at INAF Trieste and Dept. of Physics and Astronomy, University of Bologna, (44) Japanese MAGIC Group: Institute for Cosmic Ray Research (ICRR), The University of Tokyo, Kashiwa, 277-8582 Chiba, Japan), (45) Dipartimento Interateneo di Fisica dell'Universit\`a e del Politecnico di Bari, 70125 Bari, Italy, (46), Dipartimento Interateneo di Fisica dell'Universit\`a e del Politecnico di Bari, 70125 Bari, Italy, (47) Instituto de Astrof\'isica de Andaluc\'ia (CSIC), Apartado 3004, E-18080 Granada, Spain, (48) INFN Sezione di Perugia, 06123 Perugia, Italy, (49)
INAF - Istituto di Radioastronomia, Via Gobetti 101, I-40129 Bologna, Italy, (50) Department  of  Physics  and  Astronomy,  University  of  Turku,  FI-20014, Turku, Finland, (51) Universit\`{a} di Siena, I-53100, Siena, Italy\\


\appendix

\section{Details of the Multi-Wavelength analysis.}
Here we provide more detail on the observations and the analysis performed in the different energy ranges. 

\begin{table}
\caption{Observed flux of \source\ observation with MAGIC. The 95\% confidence level upper limits are also reported for observations with less than 3$\sigma$ significance.} 
\label{magic_table}
\begin{center}
\setlength{\tabcolsep}{0.15em}
\begin{tabular}{cccccc}
\hline
\multicolumn{1}{c}{MJD} &
\multicolumn{1}{c}{Time}  &
\multicolumn{1}{c}{Flux ($> 400\, \rm{GeV}$)} &
\multicolumn{1}{c}{Flux ($> 400\, \rm{GeV}$) UL} &
\multicolumn{1}{c}{Significance}\\
\multicolumn{1}{c}{} &
\multicolumn{1}{c}{(h)} &
\multicolumn{1}{c}{($\times$\,10$^{-12}$  cm$^{-2}$ s$^{-1}$)}&
\multicolumn{1}{c}{($\times$\,10$^{-12}$  cm$^{-2}$ s$^{-1}$)} &
\multicolumn{1}{c}{$\sigma$}
\\
\hline
58541.21 & 1.31 & 6.28 $\pm$ 2.16 & & 3\\
58542.27 & 1.97 & 7.42 $\pm$ 1.29 & & 7\\
58543.20 & 1.31 & - & 4.47 & 0\\
58544.19 & 0.35 & - & 1.79 & 0.7\\
58545.22 & 2.12 & 3.85 $\pm$ 1.19 & & 5\\
58547.19 & 1.04 & - & 4.65 & 1\\
\hline
\end{tabular}
\end{center}
\end{table}  

\begin{table}
\caption{Observed flux of \source\ observation with \emph{Fermi}-LAT. The exposure time for each bin is 24h, and the MJD values reported in the first column correspond to the middle of the time bin. The 95\% confidence level upper limits are reported for bins with TS < 9.}
\label{fermi_table}
\begin{center}
\setlength{\tabcolsep}{0.1em}
\begin{tabular}{cccc}
\hline
\multicolumn{1}{c}{MJD} &
\multicolumn{1}{c}{Flux ($> 300\, \textrm{MeV}$)}  &
\multicolumn{1}{c}{Flux ($> 300\, \textrm{MeV}$) UL} &
\multicolumn{1}{c}{TS}
\\
\multicolumn{1}{c}{} &
\multicolumn{1}{c}{($\times\,$10$^{-7}$  cm$^{-2}$ s$^{-1}$)} &
\multicolumn{1}{c}{($\times\,$10$^{-7}$  cm$^{-2}$ s$^{-1}$)} &
\multicolumn{1}{c}{} \\
\hline
58539.22 & 0.635 $\pm$ 0.328 & & 9.36\\
58540.22 & 0.821 $\pm$ 0.371 & & 11.75\\
58541.22 & 0.674 $\pm$ 0.316 & & 22.00\\
58542.22 & 0.549 $\pm$ 0.296 & & 11.35\\
58543.22 & - & 0.228 & 0 \\
58544.22 & 0.511 $\pm$ 0.294 & & 10.62\\
58545.22 & -  & 1.266 & 6.74\\
58546.22 & 4.664 $\pm$ 0.842 & & 138.29\\
58547.22 & - & 0.192 & 0 \\
58548.22 & - & 1.020 & 0.97\\
\hline
\end{tabular}
\end{center}

\end{table}

\begin{table}
\caption{Log and fitting results of \Xm\ observations of \source\ with $N_{\rm H}$ fixed to Galactic absorption. Fluxes are corrected for the Galactic absorption.} 
\label{xmm_table}
\begin{center}
\setlength{\tabcolsep}{0.2em}
\begin{tabular}{cccc}
\hline
\multicolumn{1}{c}{MJD} &
\multicolumn{1}{c}{Exposure time} &
\multicolumn{1}{c}{Flux ($ 3-10 \, \textrm{keV}$)}  &
\multicolumn{1}{c}{Flux ($0.3-10 \, \textrm{keV}$)}\\
\multicolumn{1}{c}{} &
\multicolumn{1}{c}{(ks)}&
\multicolumn{1}{c}{($\times\,$10$^{-12}$  erg cm$^{-2}$ s$^{-1}$)} &
\multicolumn{1}{c}{($\times\,$10$^{-12}$  erg cm$^{-2}$ s$^{-1}$)}\\
\hline
58544.10 & 25 & 2.47 $\pm$ 0.02& 8.16 $\pm$ 0.05\\
\hline
\end{tabular}
\end{center}
\end{table}

\begin{table}
\caption{Log and fitting results of \Nu\ observations of \source\  with $N_{\rm H}$ fixed to Galactic absorption. Fluxes are corrected for the Galactic absorption.} 
\label{nu_table}
\begin{center}
\setlength{\tabcolsep}{0.2em}
\begin{tabular}{cccc}
\hline
\multicolumn{1}{c}{MJD} &
\multicolumn{1}{c}{Exposure time}&
\multicolumn{1}{c}{Flux ($ 3-10 \, \textrm{keV}$)}  &
\multicolumn{1}{c}{Flux ($4-79 \, \textrm{keV}$)}\\
\multicolumn{1}{c}{} &
\multicolumn{1}{c}{(ks)}&
\multicolumn{1}{c}{($\times\,$10$^{-12}$  erg cm$^{-2}$ s$^{-1}$)} &
\multicolumn{1}{c}{($\times\,$10$^{-12}$  erg cm$^{-2}$ s$^{-1}$)}\\
\hline
58544.95 & 34 &5.69 $\pm$ 0.04 & 3.25 $\pm$ 0.02\\
\hline
\end{tabular}
\end{center}
\end{table}

\begin{table}
\caption{Log and fitting results of {\em Swift}-XRT observations of \source\  using a PL model with $N_{\rm H}$ fixed to Galactic absorption. Fluxes are corrected for the Galactic absorption.} 
\label{XRT_table}
\begin{center}
\setlength{\tabcolsep}{0.51em}
\begin{tabular}{cccc}
\hline
\multicolumn{1}{c}{MJD} &
\multicolumn{1}{c}{Exposure time} &
\multicolumn{1}{c}{Flux 0.3--10 keV}  &
\multicolumn{1}{c}{Photon index}  \\
\multicolumn{1}{c}{} &
\multicolumn{1}{c}{(s)} &
\multicolumn{1}{c}{($\times$10$^{-12}$ erg cm$^{-2}$ s$^{-1}$)} &
\multicolumn{1}{c}{($\Gamma_{\rm\,X}$)} \\
\hline
56930.87 & 1193.7  & 1.08 $\pm$ 0.20 & 2.84 $\pm$ 0.32\\
58488.34 & 229.8   & 5.37 $\pm$ 0.11 & 2.55 $\pm$ 0.30\\
58541.65 & 2462.3  & 1.69 $\pm$ 0.65 & 2.14 $\pm$ 0.51\\
58544.12 & 536.6   & 8.16 $\pm$ 0.81 & 2.50 $\pm$ 0.14\\
58544.84 & 401.2   & 1.78 $\pm$ 0.21 & 2.00 $\pm$ 0.15\\
58544.97 & 529.0   & 1.75 $\pm$ 0.19 & 2.29 $\pm$ 0.15\\
58547.43 & 2497.3  & 3.83 $\pm$ 0.23 & 2.73 $\pm$ 0.10\\
58551.62 & 1648.2  & 2.73 $\pm$ 0.26 & 2.89 $\pm$ 0.14\\
58554.60 & 2582.2  & 5.24 $\pm$ 0.29 & 2.61 $\pm$ 0.92 \\
58560.71 & 2467.3  & 2.38 $\pm$ 0.17 & 2.88 $\pm$ 0.13 \\          
\hline
\end{tabular}
\end{center}
\end{table}  

\begin{table}
\caption{Observed magnitudes with \emph{XMM-OM}.}
\label{om_table}
\begin{center}
\setlength{\tabcolsep}{0.2em}
\begin{tabular}{cccc}
\hline
\multicolumn{1}{c}{MJD} &
\multicolumn{1}{c}{B}  &
\multicolumn{1}{c}{U} & 
\multicolumn{1}{c}{W1}\\
\multicolumn{1}{c}{} &
\multicolumn{1}{c}{(AB mag)} &
\multicolumn{1}{c}{(AB mag)} &
\multicolumn{1}{c}{(AB mag)}\\
\hline
58544.10 & $16.639\pm0.005$ &$17.089\pm0.008$& $17.406\pm0.009$ \\ 
\hline
\end{tabular}
\end{center}
\end{table}

\begin{table}
\caption{Observed magnitudes for \source\ with Siena Observatory.}
\label{siena_table}
\begin{center}
\setlength{\tabcolsep}{0.2em}
\begin{tabular}{ccc}
\hline
\multicolumn{1}{c}{MJD} &
\multicolumn{1}{c}{Observed magnitude}\\
\multicolumn{1}{c}{} &
\multicolumn{1}{c}{(mag)}\\
\hline
58542.17 & $15.43 \pm 0.02$ \\
58551.14 & $15.50 \pm 0.02$ \\
58564.10 & $15.56 \pm 0.03$ \\
\hline
\end{tabular}
\end{center}
\end{table}

\begin{table}
\caption{Observed magnitudes with KVA.}
\label{kva_table}
\begin{center}
\setlength{\tabcolsep}{0.2em}
\begin{tabular}{ccc}
\hline
\multicolumn{1}{c}{MJD} &
\multicolumn{1}{c}{Observed magnitude}  \\
\multicolumn{1}{c}{} &
\multicolumn{1}{c}{(mag)}\\
\hline
58541.22  & $ 15.46  \pm  0.02 $ \\
58542.21  & $ 15.42  \pm  0.02 $ \\
58543.23  & $ 15.5  \pm  0.02 $ \\
58544.24  & $ 15.43  \pm  0.06 $ \\
58545.24  & $ 15.44  \pm  0.02 $ \\
58547.23  & $ 15.39  \pm  0.02 $ \\
58549.24  & $ 15.45  \pm  0.02 $ \\
58551.22  & $ 15.48  \pm  0.02 $ \\
58552.24  & $ 15.54  \pm  0.02 $ \\
58554.21  & $ 15.48  \pm  0.02 $ \\
58555.2  & $ 15.39  \pm  0.02 $ \\
58556.2  & $ 15.46  \pm  0.02 $ \\
58557.19  & $ 15.51  \pm  0.02 $ \\
58558.17  & $ 15.52  \pm  0.02 $ \\
58559.24  & $ 15.57  \pm  0.02 $ \\
58560.22  & $ 15.66  \pm  0.08 $ \\
58565.25  & $ 15.68  \pm  0.03 $ \\
58580.18  & $ 15.78  \pm  0.02 $ \\
58584.2  & $ 15.8  \pm  0.03 $ \\
58586.19  & $ 15.76  \pm  0.02 $ \\
58596.14  & $ 15.99  \pm  0.04 $ \\
58600.16  & $ 15.92  \pm  0.03 $ \\
58603.14  & $ 15.93  \pm  0.03 $ \\
58609.95  & $ 16.0  \pm  0.06 $ \\
58617.12  & $ 16.05  \pm  0.04 $ \\
58633.9  & $ 15.89  \pm  0.03 $ \\
58643.93  & $ 16.11  \pm  0.03 $ \\
58658.97  & $ 16.14  \pm  0.03 $ \\
58666.94  & $ 16.14  \pm  0.03 $ \\
58673.92  & $ 16.0  \pm  0.03 $ \\
58678.91  & $ 16.08  \pm  0.04 $ \\
58687.9  & $ 15.95  \pm  0.03 $ \\
58692.91  & $ 15.98  \pm  0.03 $ \\
58699.92  & $ 16.03  \pm  0.03 $ \\
58706.89  & $ 16.0  \pm  0.04 $ \\
58714.88  & $ 15.93  \pm  0.03 $ \\
58718.87  & $ 15.86  \pm  0.03 $ \\
\hline
\end{tabular}
\end{center}
\end{table}

\begin{table*}
\caption{Observed magnitudes for \source\ obtained with {\em Swift}-UVOT.}
\label{uvot_table}
\setlength{\tabcolsep}{0.41em}
\begin{center}
\begin{tabular}{ccccccc}
\hline
\multicolumn{1}{c}{MJD}  &
\multicolumn{1}{c}{V}    &
\multicolumn{1}{c}{B}    &
\multicolumn{1}{c}{U}    &
\multicolumn{1}{c}{W1}   &
\multicolumn{1}{c}{M2}   &
\multicolumn{1}{c}{W2}   \\
\multicolumn{1}{c}{} &
\multicolumn{1}{c}{(mag)} &
\multicolumn{1}{c}{(mag)} &
\multicolumn{1}{c}{(mag)} &
\multicolumn{1}{c}{(mag)} &
\multicolumn{1}{c}{(mag)} &
\multicolumn{1}{c}{(mag)}  \\
\hline
56930 & 16.74 $\pm$ 0.14 & 17.72 $\pm$ 0.16 &  16.96 $\pm$ 0.12 & 17.50 $\pm$ 0.16  & 17.71 $\pm$ 0.14 & 17.74 $\pm$ 0.12 \\   
58488 & --               & --               &  16.58 $\pm$ 0.07 & --                & --               & --               \\
58541 & 15.95 $\pm$ 0.06 & 16.56 $\pm$ 0.06 &  15.97 $\pm$ 0.06 & 16.19 $\pm$ 0.07  & 16.60 $\pm$ 0.09 & 16.52 $\pm$ 0.07 \\
58544 & 15.89 $\pm$ 0.07 & 16.60 $\pm$ 0.06 &  15.92 $\pm$ 0.07 & 16.23 $\pm$ 0.08  & 16.67 $\pm$ 0.10 & 16.58 $\pm$ 0.07 \\
58547 & 15.87 $\pm$ 0.05 & 16.49 $\pm$ 0.05 &  15.88 $\pm$ 0.06 & 16.25 $\pm$ 0.07  & 16.58 $\pm$ 0.08 & 16.60 $\pm$ 0.07 \\
58551 & 16.07 $\pm$ 0.07 & 16.59 $\pm$ 0.06 &  16.01 $\pm$ 0.07 & 16.36 $\pm$ 0.08  & 16.70 $\pm$ 0.09 & 16.63 $\pm$ 0.08 \\
58554 & 15.87 $\pm$ 0.05 & 16.52 $\pm$ 0.05 &  15.82 $\pm$ 0.06 & 16.24 $\pm$ 0.07  & 16.50 $\pm$ 0.08 & 16.52 $\pm$ 0.07 \\
58560 & 16.09 $\pm$ 0.06 & 16.77 $\pm$ 0.06 &  16.10 $\pm$ 0.06 & 16.37 $\pm$ 0.08  & 16.70 $\pm$ 0.08 & 16.84 $\pm$ 0.07 \\              
\hline
\end{tabular}
\end{center}
\end{table*}


\bsp	
\label{lastpage}
\end{document}